\documentclass[
 reprint,
superscriptaddress,
nofootinbib,
 amsmath,amssymb,
 aps,
 prl,
floatfix,
]{revtex4-1}

\usepackage{graphicx,color}
\usepackage{dcolumn}
\usepackage{bm}
\usepackage{hyperref}
\usepackage{physics}
\usepackage{soul}
\usepackage{cancel}
\usepackage{mathtools}
\usepackage{cprotect}
\usepackage{enumitem}
\usepackage[normalem]{ulem}

\usepackage[multidot]{grffile}

\usepackage[caption=false]{subfig}

\usepackage[dvipsnames]{xcolor}
\definecolor{C0}{RGB}{31, 119, 180}
\definecolor{C1}{RGB}{255, 127, 14}
\definecolor{C2}{RGB}{44, 160, 44}
\definecolor{C3}{RGB}{214, 39, 40}
\definecolor{C4}{RGB}{148, 103, 189}
\definecolor{C5}{RGB}{140, 86, 75}

\begin{document}

\title{Masking equation of state effects in binary neutron star mergers}

\author{Antonios Tsokaros}
\affiliation{%
Department of Physics, University of Illinois Urbana-Champaign, Urbana, IL 61801, USA
}%
\affiliation{National Center for Supercomputing Applications, University of Illinois Urbana-Champaign, Urbana, IL 61801, USA}
\affiliation{Research Center for Astronomy and Applied Mathematics, Academy of Athens, Athens 11527, Greece}
\author{Jamie Bamber}
\affiliation{%
Department of Physics, University of Illinois Urbana-Champaign, Urbana, IL 61801, USA
}
\author{Milton Ruiz}%
\affiliation{%
 Departament d’Astronomia i Astrof\'{i}sica, Universitat de Val\`{e}ncia, C/ Dr Moliner 50, 46100, Burjassot (Val\`{e}ncia), Spain
}
\author{Stuart L. Shapiro}
\affiliation{%
Department of Physics, University of Illinois Urbana-Champaign, Urbana, IL 61801, USA
}%
\affiliation{
Department of Astronomy, University of Illinois Urbana-Champaign, Urbana, IL 61801, USA
}
\affiliation{National Center for Supercomputing Applications, University of Illinois Urbana-Champaign, Urbana, IL 61801, USA}

\date{\today}

\begin{abstract} 
Recent nonmagnetized studies of binary neutron star mergers have indicated the possibility of identifying equation of state features, such as a phase transition or a quark-hadron crossover, based on the frequency shift of the main peak in the postmerger gravitational wave spectrum. By performing a series of general relativistic, magnetohydrodynamic simulations we show that similar frequency shifts can be obtained due to the effect of the magnetic field. The existing degeneracy can either mask or nullify a shift due to a specific equation of state feature, and therefore the interpretation of observational data is more complicated than previously thought, requiring a more complete treatment that would necessarily include the neutron star's magnetic field.
\end{abstract}

\maketitle

\textit{Introduction.}\textemdash
One of the premises of multimessenger astronomy is a better understanding of the nature of matter at supranuclear densities.
In event GW170817~\cite{LIGOScientific:2017vwq}, although the postmerger phase was not observed, 
constraints on the neutron star (NS) radii and EOS were placed \cite{PhysRevLett.121.161101}, that resulted 
to a prediction that a $\sim 1.4 M_\odot$ NS will have a radius of $\sim 11.9$ km.
At the same time, X-ray observations from NASA's NICER and XMM-Newton instruments were able to measure
\cite{Miller:2019cac,Riley:2019yda,Riley:2021pdl,Miller_2021} 
the masses and radii of two millisecond pulsars. PSR J0030+0451 had a mass of $\sim 1.4 M_\odot$ (canonical NS mass), and 
radius $\sim 13$ km, while PSR J0740+6620, had mass $2.1 M_\odot$ and radius $\sim 13.7$ km.
Constraints such as the above help us understand the cold (zero-temperature) equation of state (EOS) of a NS. 
With the advent of next generation gravitational wave (GW) detectors, such as the Cosmic Explorer \cite{Abbott_2017} 
or the Einstein Telescope \cite{Maggiore_2020},
we will be able to observe the postmerger GW signal which is very important since the densities after merger
can reach $\sim 6 n_0$, where $n_0=0.16\ {\rm fm^{-3}}$ the nuclear saturation number density, while the temperatures can be as high 
as $100$ MeV~\cite{Sekiguchi:2011mc}. This will enable us to understand the nature of matter at the most extreme limits.

In the past years, numerical relativistic simulations helped us understand the high density postmerger phase, through the oscillation
frequencies of the GW signal 
\cite{Hotokezaka:2011dh,
Bauswein:2011tp,
Takami:2014zpa,
Bernuzzi2015,
Bauswein:2015yca,
Rezzolla:2016nxn,
Maione_2016,
Vretinaris:2019spn,
Soultanis:2021oia}.
In a typical NS merger scenario where the remnant does not immediately 
collapse to a black hole, there are three main frequencies associated with the GW spectrum. In this work we will only focus on 
the most prominent peak denoted by $f_2$ (sometimes by $f_{\textup{peak}}$), which can be detected by advanced GW detectors and which 
corresponds to the fundamental quadrupolar $l=m=2$ fluid mode \cite{Stergioulas:2011gd}.

How and when frequency $f_2$ changes with respect to an EOS has led a number of authors to place strict constraints
on the latter, which can have important consequences on the composition of matter in NSs. 
For example, in \cite{Bauswein:2018bma,Blacker:2020nlq}
the authors found a significant deviation from an empirical relation between $f_2$ and the tidal deformability if a 
strong first-order phase transition (PT) leads to the formation of a gravitationally stable, extended quark matter core in 
the postmerger remnant. Under such conditions $f_2$ was found to {\it increase} (relative to its value in the absence of a PT), which can be used to detect a PT using future GW detections. An alternative scenario was proposed in \cite{Weih:2019xvw} where the postmerger signal exhibits two distinct $f_2$ frequencies, one before and one after the PT. The second fundamental frequency was {\it larger} than the first one, and corresponds to a significant quark core.

The common thread in the appearence of quark matter is the decrease of pressure support, which is accompanied by a decrease
in radius or an increase of compactness, which in turn leads to a {\it higher} GW frequency emitted from the remnant.
An alternative to a PT is a continuous crossover from hadronic to quark matter \cite{Huang:2022mqp}. Quark-hadron-crossover (QHC) EOSs 
\cite{Masuda2013,Baym2018,Baym:2019iky,Kojo2022} exhibit a peak in the sound speed \cite{PhysRevC.109.065803}
which can be taken as a signature
for the onset of quark matter formation. A pronounced peak in the sound speed, (stiff QHC) leads to a {\it lower} $f_2$
for a range of remnant masses, which in turn can be used to distigush such EOSs from those that have a PT and lead to 
{\it higher} $f_2$ frequencies. On the other hand, a less pronounced peak in the sound speed (soft QHC) can have a
{\it lower} or {\it higher} $f_2$ depending on the mass of the remnant \cite{Huang:2022mqp}.

The shift of the $f_2$ frequency in all of the above works is computed through hydrodynamic (nonmagnetized) simulations. 
Given the fact that NSs have large magnetic fields that reach values beyond magnetar strength during merger 
\cite{Kiuchi:2014hja,Aguilera-Miret:2023qih}, we go one step further in this work by assessing for the first time 
the influence of the magnetic field on the postmerger
frequency $f_2$, and discuss the ramifications with respect to the possibility of distinguishing one EOS versus another.
We show that for GW shifts $\lesssim 200$ Hz, the magnetic field can mimic analogous shifts 
coming from the EOS sector and therefore either nullify possible EOS effects, or masquerade as them. 

\textit{Simulations.}\textemdash
We perform general relativistic magnetohydrodynamic simulations using two EOS (SLy \cite{Douchin:2001sv} and ALF2 \cite{Alford:2004pf}),
with two different Arnowitt–Deser–Misner (ADM) masses ($2.57\ M_\odot$ and $2.7\ M_\odot$) and three topologies of magnetic fields: 
i) A pulsar-like magnetic field, i.e. poloidal dipolar topology that extends from the interior of the NS to the exterior. We employ four
different magnetic field strengths with average values (within the NS) of 
$7.9\times 10^{14}$ G, $2.0\times 10^{15}$ G, $7.9\times 10^{15}$ G, and $3.2\times 10^{16}$ G;
ii) an interior-only toroidal magnetic field with average value $2.2\times 10^{16}$ G; and 
iii) an interior-only poloidal magnetic field with average value $2.7\times 10^{16}$ G.
The choice of EOSs although arbitrary are in broad agreement with current observations and are sufficient for the proof-of-principle
scenario proposed here. The values of the ADM masses of the binary are chosen so that both supramassive \cite{Cook92b}
and hypermassive \cite{Baumgarte:1999cq} remnants are formed, and a clear identification of the fundamental frequency $f_2$ can be made.
The range of values for the magnetic field are typical in numerical investigations 
\cite{Siegel:2013nrw,Ciolfi:2019fie,Ruiz:2020via,Ruiz:2021qmm,Bamber:2024kfb,Kiuchi:2023obe},
and dictated by the highest resolution that we can afford, which is $\Delta x_{\rm min}\approx 90$ m,
in order to ensure that magnetic instabilities will be resolved (see \cite{Bamber2024un}). 
In comparison the hydrodynamic simulations in \cite{Huang:2022mqp} employ a resolution of $185$ m (in their highest resolution), 
while the ones in \cite{Weih:2019xvw} have $237$ m.
As shown in \cite{Kiuchi:2014hja,Aguilera-Miret:2023qih} the magnetic field is amplified in a short timescale after the merger 
of the two NSs and leads to extremely high field strengths ($ > 10^{16}$ G). To account for thermal effects during the collision
of the two NSs we add an ideal gas pressure component to the cold EOS (as assumed in the initial quasiequilibrium binary configuration).
We employ an ideal-gas index $\Gamma_{\rm th}=5/3$, although a higher value (but $\leq 2$) may lead to possible higher power
in the postmerger oscillation modes and even more pronounced frequency shifts. 
Details on the numerical setup can be found in \cite{Bamber2024un}, where the same magnetic fields as here are expressed
in terms of their maximum instead of average values.

\textit{Results and discussion.}\textemdash
It is well known that following merger the existence of a magnetic field has important consequences on the merger remnant that can distinguish its evolution from the nonmagnetized  analogue. As the two NSs come in contact, a velocity-shear layer is formed at the contact interface which is subject to the
Kelvin-Helmholtz (KH) instability.  Small vortices are then created which wind
up the magnetic field in a timescale of $\sim 0.01$ ms (much shorter than the
dynamical timescale). This  leads to an exponential amplification of the magnetic
field which reaches values $\gtrsim 10^{15}\;\rm G$ or even larger in less than
$\sim 5$ ms~\cite{Price:2006fi,Kiuchi:2014hja,Kiuchi:2015sga,Aguilera-Miret:2020dhz}.
In addition to the KH instability other mechanisms, such as the
magnetorotational instability (MRI) \cite{Balbus:1991ay}, and magnetic
winding \cite{Shapiro:2000zh,Duez:2006qe}, help to amplify the magnetic field. This powerful magnetic
field will play a crucial role (in addition to the total mass of the system) in
determining a number of electromagnetic observables, and, ultimately, the lifetime of the
remnant itself.

\begin{figure}
    \centering
    \includegraphics[width=\linewidth]{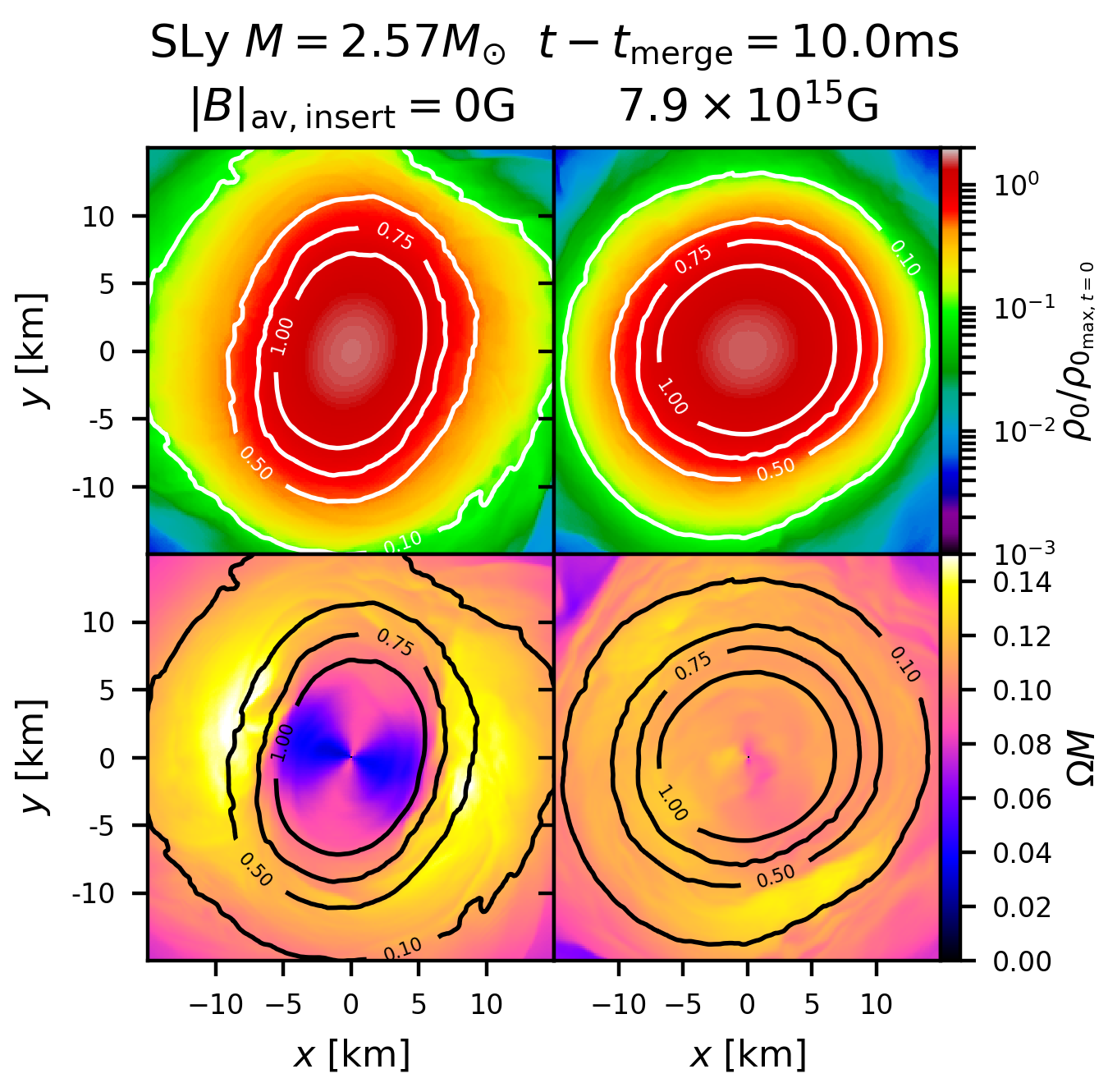}
    \caption{Rest-mass density (top panels) and angular velocity (bottom panels) on the equatorial plane
for the nonmagnetized SLy $2.57\ M_\odot$ case (left column) and the $7.9\times 10^{15}$ G pulsar-like case (right column),
$10$ ms after merger. White and black lines signify contour plots of density multiples of $\rho_{0\,\rm max,t=0}$.}
     \label{fig:rho_omega}
\end{figure}

\begin{figure*}
    \centering
    \includegraphics[width=\linewidth]{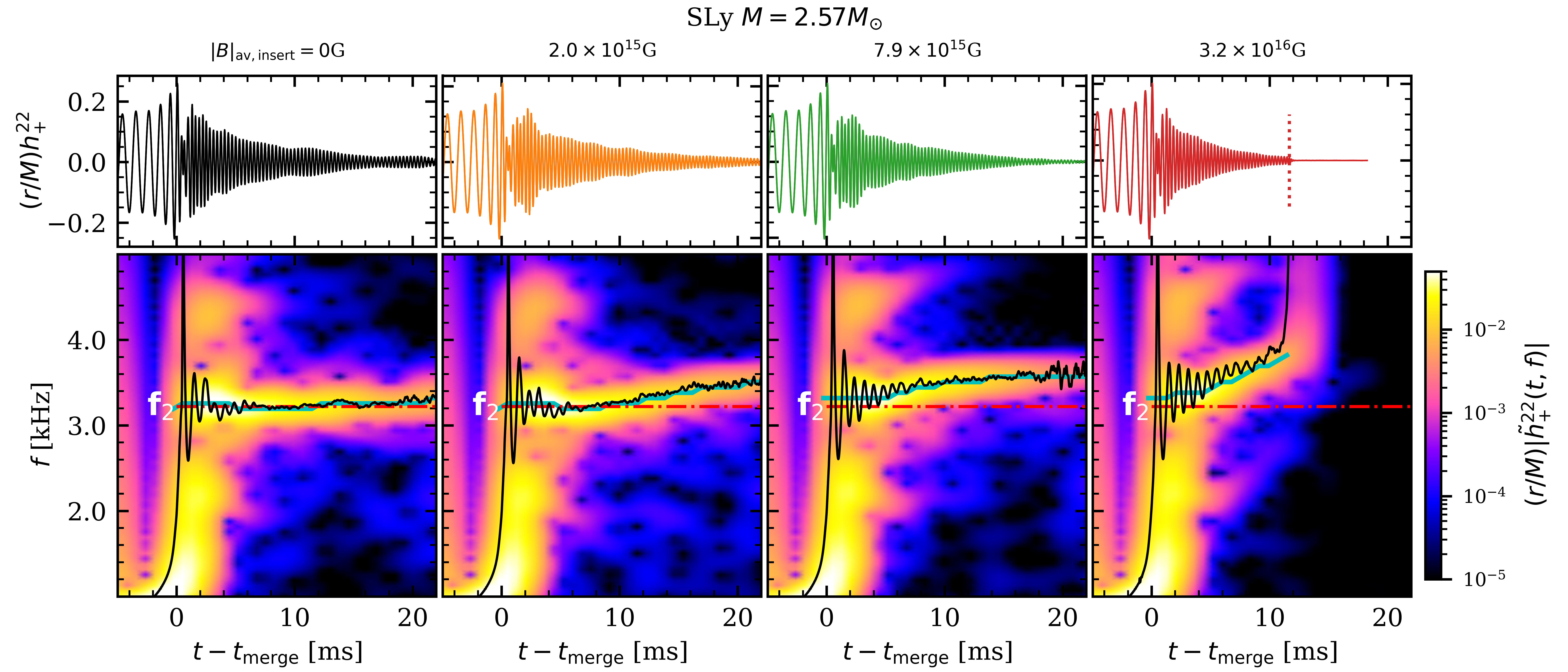}
    \caption{Strain $h^{22}_{+}$ (top) and spectrograms (bottom) for the SLy $M=2.57\ M_\odot$ case for the nonmagnetized
(left column) and magnetized with a pulsar-like magnetic field (columns 2-4). In the bottom panels the cyan curve shows the 
instantaneous $f_2$ frequency which corresponds to the maximum of the spectrograms, while the black line shows the instantaneous GW frequency $f_{\textup{GW}}$ defined by $f_{\textup{GW}} := \vert \dd \phi_{22} / \dd t \vert$ where $\phi_{22}$ is the phase of the $22$ mode given by $\phi_{22} = \textup{arg}(h_{+} - i h_{\cross})$. The dashed-dotted lines show the 
nomagnetized value of $f_2$.}
    \label{fig:SLy_spectrograms}
\end{figure*}

In Fig. \ref{fig:rho_omega} we plot the rest-mass density (upper pannels) and the fluid angular velocity
(bottom panels) for two representative cases, $10$ ms after merger: The nonmagnetized SLy case (left column) 
and the magnetized pulsar-like $7.9\times 10^{15}$ G case (right column). The most obvious difference between them
is the fluid angular velocity for densities larger than
$\rho_{0\,\rm max,t=0}$ (the maximum density of the quasiequilibrium configurations). While in the nonmagnetized
binary the angular velocity profile exhibits a depression at the origin \cite{Kastaun:2014fna,Hanauske:2016gia}
and a clear maximum at $0.5 \rho_{0\,\rm max,t=0}$,
in the magnetized case we have a more or less uniform angular velocity core whose maximum value is less
than the nonmagnetized one \cite{Ruiz:2021qmm}. At the same time, from the density contour lines we observe that the magnetized
remnant is more axisymmetric than the nonmagnetized one, which even 10 ms after merger still preserves 
its ellipsoidal shape.  

These preliminary hydrodynamic differences manifest in the oscillations spectra of the GW as seen in Fig. \ref{fig:SLy_spectrograms}. In this representative plot we show in the top row the $\ell=m=2$ GW strain for the $+$ polarization, while in the bottom row we show the corresponding spectograms along with the instantaneous maximum of the power spectral density (cyan line) that corresponds to the $f_2$ frequency. The panels correspond to the SLy $M=2.57\ M_\odot$  simulations with the nonmagnetized case in the left column, and the three pulsar-like magnetized cases in the rest (the highest magnetic field in right column). The remnants in all cases are supramassive \cite{Cook92b} NSs that should persist for a long timescale. This turns out to be true for all cases, apart for the one with the strongest magnetic field that collapses to a black hole approximately 12 ms after merger (vertical dashed line on upper right panel). The reason for this unexpected outcome is because due to the strong magnetic field the remnant, which by that time is almost uniformly rotating, has lost sufficient angular momemntum and cannot support its mass. In the rest-mass-angular momentum ($M_0-J$) parameter space, the remnant in the far right column finds itself in the unstable region and thus collapses to a black hole (details are provided in \cite{Bamber2024un}).

\begin{figure}
    \centering
    \includegraphics[width=\linewidth]{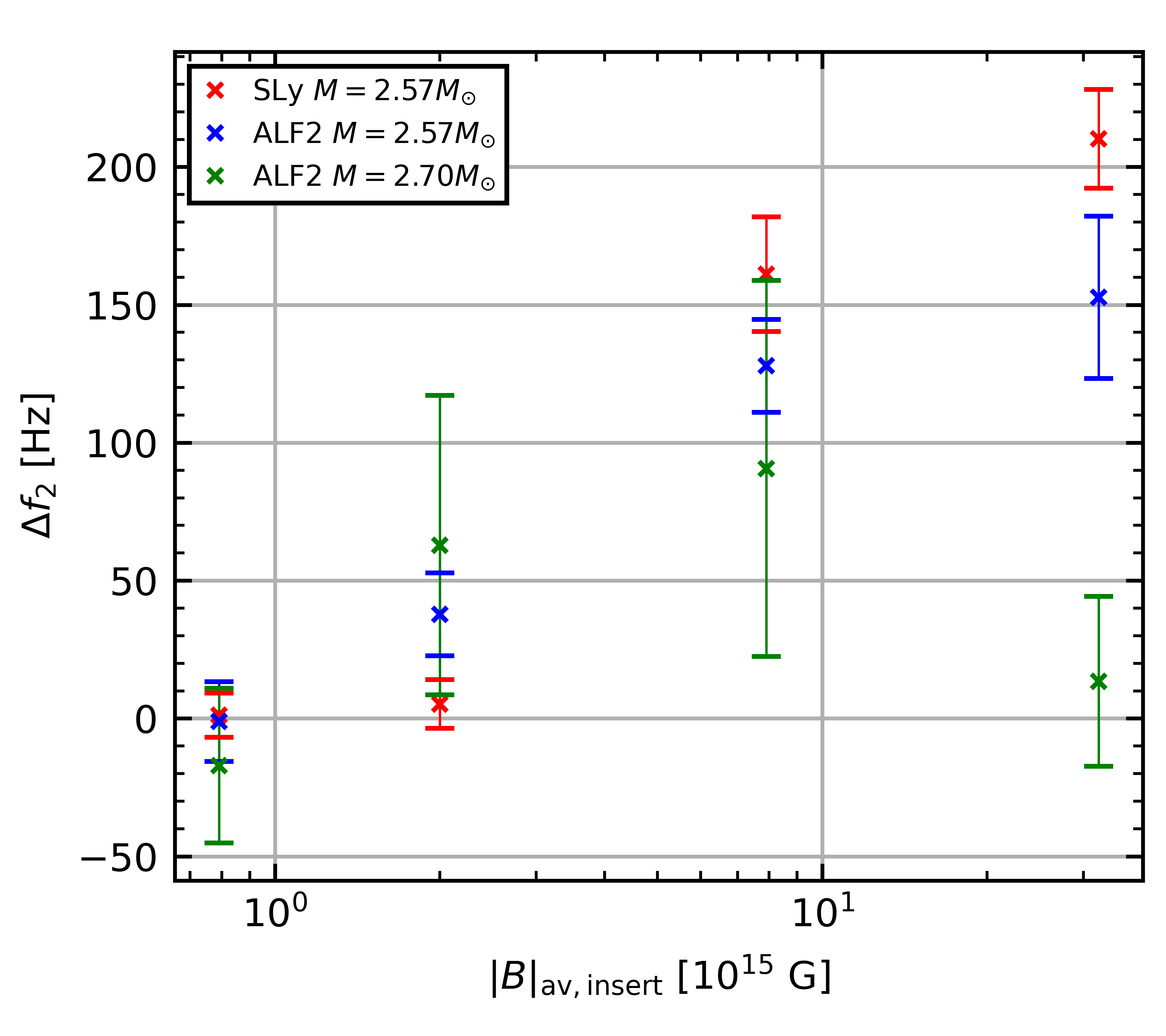}
   \caption{Frequency shift $\Delta f_2=f_2^{\rm mag} - f_2^{\rm hyd}$ with respect to the nonmagnetized case.}
    \label{fig:f2}
\end{figure}

It is apparent from the spectrograms in columns 2-4 of Fig. \ref{fig:SLy_spectrograms} that the quadrupole frequency $f_2$ in the magnetized cases exhibits a shift to {higher} frequencies with respect to the nonmagnetized case. For the latter, the behavior is roughly consistent with that reported in \cite{Rezzolla:2016nxn}: an initial transient regime up to $\sim 3$ms postmerger where the instantaneous dominant frequency is slightly higher, followed by a quasistationary regime where the frequency is constant. For the magnetized cases the shift is clearly depicted in Fig. \ref{fig:f2} for the two EOS and two masses with the pulsar-like magnetic field that we investigated. This frequency is obtained by fitting Gaussians to the corresponding amplitude spectral density ($2 f^{1/2} \vert \hat{h}(f) \vert$) peaks. The error bars show the $2\sigma$ frequency spread across the four different Tukey window function parameters  ($0.01, 0.05, 0.1, 0.25$) used here. The largest frequency shift happens not surprisingly for the largest magnetic field and the smaller mass remnant ($\sim 200$ Hz), but even smaller magnetic fields can produce shifts $\sim 150$ Hz as in the pulsar-like SLy $M=2.57\ M_\odot$ case with average magnetic field $7.9\times 10^{15}$ G. These magnetar strength magnetic fields are commonplace in binary mergers \cite{Kiuchi:2015sga} which predict that maximum magnetic fields of the order of $\sim 10^{17}$ G can be realized.

Let us now recall that $f_2$ frequency shifts in a merger remnant can be caused by a number of reasons: \begin{enumerate}[topsep=0pt,itemsep=-1ex,partopsep=1ex,parsep=1ex]
\item The existence of a PT as shown in \cite{Bauswein:2018bma,Most:2018eaw,Blacker:2020nlq} leads to a {\it higher} frequency with a shift in $f_2$ of up to $\sim 500$ Hz. More generally, the presence of anomalous, non-convex dynamics \cite{Rivieccio:2024sfm} may be responsible for such large shifts. Similarly for the scenario in \cite{Weih:2019xvw}.
\item The existence of a QHC EOS \cite{Huang:2022mqp} can lead to a {\it higher} or {\it lower} frequency, with a shift of up to $\sim 100$ Hz, depending
on various parameters including the mass of the system.
\item Finite temperature effects can lead to a {\it higher} postmerger frequency \cite{Raithel:2023gct,Fields2023,Villa-Ortega:2023cps} up to $\sim 200$ Hz.
\item The stiffness of the EOS (mass-radius positive slope) leads to a {\it higher} frequency \cite{Raithel:2022orm} with an shift of up to $\sim 600$ Hz.
\item Out-of-equilibrium effects, such as bulk viscosity \cite{Chabanov:2023blf,Most_2024}, can produce frequency shifts up to $\sim 300$ Hz.
\item The magnetic field \cite{Ruiz:2021qmm,Bamber2024un} leads to a {\it higher} frequency with an increase of up to $\sim 200$ Hz.
\item The spin of the prior NSs can also lead to a {\it higher} (for antialigned) or  {\it lower} (for aligned) postmerger frequency \cite{East:2019lbk} with a shift of up to $\sim 200$ Hz.
\end{enumerate}
Apart from the last two items in the list above, everything else is related to the as-yet unknown EOS, either in its cold or in its hot sector. The magnitude of the predicted shifts varies for each of the reasons above, but the overlap is significant. In particular this means that, any of the shifts predicted by items 1-5 can be {\it masked by the magnetic field} (or even the prior NS spin), and therefore any interpretation of observational data should be done with caution. 
Extensions to general relativity can also modify the postmerger frequency in multiple ways. For example, 
massive Damour-Esposito-Farese-type scalar-tensor theory has been found to produce {\it higher} or {\it lower} $f_2$ frequencies depending on the EOS \cite{Lam:2024azd}.

\begin{figure}
    \centering
    \includegraphics[width=\linewidth]{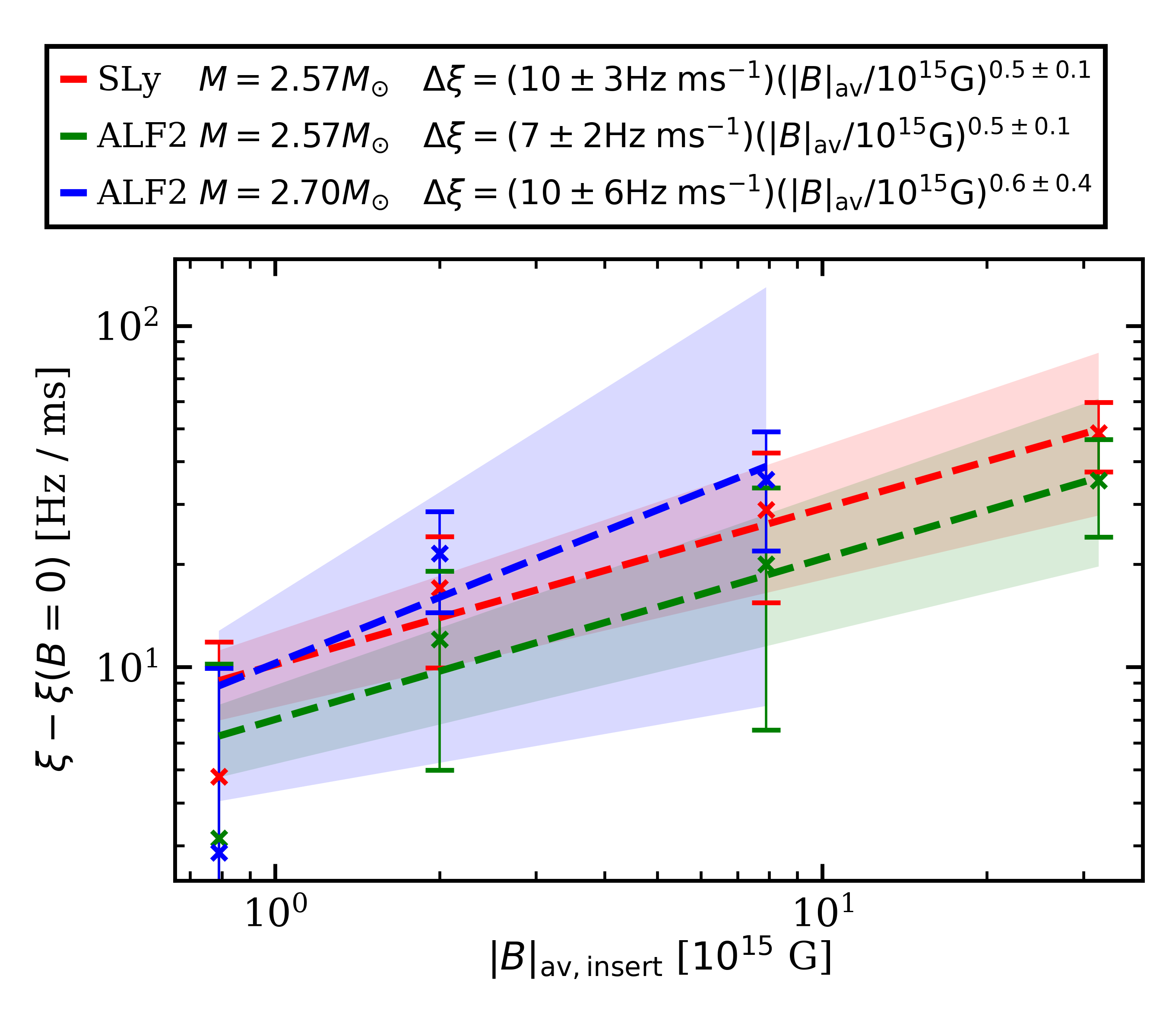}
    \caption{Growth rate for the frequency $f_2$ as a function of the initial average magnetic field strength. } 
\label{fig:drift_vs_B}
\end{figure}

Another feature evident from the spectrograms in Fig. \ref{fig:SLy_spectrograms} is that in the presence of a magnetic field, the dominant frequency, which initially is consistent with the nonmagnetized case, subsequently drifts to higher frequencies. The stronger the magnetic field the faster the rate of drift. We can attribute this behavior to the loss of angular momentum support and the remnant becoming more compact (the dominant frequency roughly scales with the square root mean density $\sim \sqrt{M_{\rm remnant}/R_{\rm remnant}^3}$ \cite{Andersson:1997rn,Chakravarti:2019sdc}). The fact that the $f_2$ frequency increases with time as the remnant becomes more compact has been found in nonmagnetized studies \cite{Sekiguchi2011,Hotokezaka2013,Maione_2016,Radice:2016rys} as a result of the loss of angular momentum through GW emission and/or a transition to a denser phase of nuclear matter. Once again, we emphasize here that the magnetic field can also be responsible for such behavior and therefore the existing degeneracy worsens.

To quantify the time-dependent shift we fit a linear growth rate to the instantaneous $f_2$ frequency as 
\begin{equation}
    f_2(t) = f_{2,0} + \xi \cdot t, 
\end{equation}
where $f_{2,0},\xi$ are constants. If we focus on the cases with initial
pulsar-like magnetic fields and exclude the ALF2 $M=2.70M_{\odot}$, 
$\vert B \vert_{\textup{av},\textup{insert}} = 3.2\times 10^{16}$G outlier we obtain
the results in Fig. \ref{fig:drift_vs_B}. The growth rate can be approximated with a power law relationship
\begin{equation}
    \xi - \xi(B=0) \approx \left(10 \pm 2\; \textup{Hz}\;\textup{ms}^{-1}\right)\left[\frac{\vert B \vert_{\textup{av},\textup{insert}}}{10^{15}\textup{G}}\right]^{0.5 \pm 0.2}
\end{equation}
which unavoidably would need further studies with more binary masses, EOSs, and low prior B-fields
to establish any universality. 


\textit{Conclusions.}\textemdash
The authors in \cite{Chatziioannou2017} conclude that the frequency $f_2$ can be measured to within about 36 Hz, 27 Hz, and 45 Hz at $90\%$ confidence level for a stiff, moderate, and soft EOS respectively for a post-merger signal-to-noise ratio of 5. Therefore, although a frequency shift due to a magnetic field is in principle detectable, the degeneracy problem (items 1-7 above) is severe.
In \cite{Bauswein:2018bma} it is accurately stated that ``To identify clear evidence for a phase transition it is indispensable to
ensure that a particular signature can only be caused by the presence of a phase transition''. 
A magnetic field can produce a $f_2$ frequency shift as large as an EOS feature and therefore the aforementioned desirable uniqueness
breaks down. In cases with a QHC EOS and a {\it lower} frequency prediction 
\cite{Huang:2022mqp}, it is conceivable that this effect can be completely masked by an analogous {\it increase} of the frequency
due to the magnetic field. In summary, what we argue here is that the premise of identifying a magnetic field signature or an EOS feature (such as a PT or a crossover) based on the $f_2$ frequency shift is not as straightforward as previously thought, since degeneracies can either mask or nullify an observation. An analysis that goes beyond the simple identification of $f_2$ shifts, along the lines of \cite{Bauswein:2014qla} can be promissing, although a full Bayesian analysis \cite{Radice:2016rys} using waveform templates that include temperature-dependent EOS, magnetic fields, and neutrino radiative transport would be ideal. The impact of the magnetic field topology on the $f_2$ frequency shift is another open question. We plan to investigate these scenarios in future studies.

\textit{Acknowledgments.}\textemdash
This work was supported in part by National Science Foundation (NSF) Grants No.
PHY-2308242, No. OAC-2310548 and No. PHY-2006066 to the University of Illinois
at Urbana-Champaign. A.T. acknowledges support from the National Center for
Supercomputing Applications (NCSA) at the University of Illinois at
Urbana-Champaign through the NCSA Fellows program.  M.R. acknowledges support
by the Generalitat Valenciana Grant CIDEGENT/2021/046 and by the Spanish
Agencia Estatal de Investigaci\'on (Grant PID2021-125485NB-C21).   This work
used Stampede2 at TACC and Anvil at Purdue University through allocation
MCA99S008, from the Advanced Cyberinfrastructure Coordination Ecosystem:
Services \& Support (ACCESS) program, which is supported by National Science
Foundation grants \#2138259, \#2138286, \#2138307, \#2137603, and \#2138296.
This research also used Frontera at TACC through allocation AST20025. Frontera
is made possible by NSF award OAC-1818253.  The authors thankfully acknowledge
the computer resources at MareNostrum and the technical support provided by the
Barcelona Supercomputing Center (AECT-2023-1-0006).

\bibliography{apssamp}

\providecommand{\noopsort}[1]{}\providecommand{\singleletter}[1]{#1}%
\begin{thebibliography}{66}%
\makeatletter
\providecommand \@ifxundefined [1]{%
 \@ifx{#1\undefined}
}%
\providecommand \@ifnum [1]{%
 \ifnum #1\expandafter \@firstoftwo
 \else \expandafter \@secondoftwo
 \fi
}%
\providecommand \@ifx [1]{%
 \ifx #1\expandafter \@firstoftwo
 \else \expandafter \@secondoftwo
 \fi
}%
\providecommand \natexlab [1]{#1}%
\providecommand \enquote  [1]{``#1''}%
\providecommand \bibnamefont  [1]{#1}%
\providecommand \bibfnamefont [1]{#1}%
\providecommand \citenamefont [1]{#1}%
\providecommand \href@noop [0]{\@secondoftwo}%
\providecommand \href [0]{\begingroup \@sanitize@url \@href}%
\providecommand \@href[1]{\@@startlink{#1}\@@href}%
\providecommand \@@href[1]{\endgroup#1\@@endlink}%
\providecommand \@sanitize@url [0]{\catcode `\\12\catcode `\$12\catcode `\&12\catcode `\#12\catcode `\^12\catcode `\_12\catcode `\%12\relax}%
\providecommand \@@startlink[1]{}%
\providecommand \@@endlink[0]{}%
\providecommand \url  [0]{\begingroup\@sanitize@url \@url }%
\providecommand \@url [1]{\endgroup\@href {#1}{\urlprefix }}%
\providecommand \urlprefix  [0]{URL }%
\providecommand \Eprint [0]{\href }%
\providecommand \doibase [0]{http://dx.doi.org/}%
\providecommand \selectlanguage [0]{\@gobble}%
\providecommand \bibinfo  [0]{\@secondoftwo}%
\providecommand \bibfield  [0]{\@secondoftwo}%
\providecommand \translation [1]{[#1]}%
\providecommand \BibitemOpen [0]{}%
\providecommand \bibitemStop [0]{}%
\providecommand \bibitemNoStop [0]{.\EOS\space}%
\providecommand \EOS [0]{\spacefactor3000\relax}%
\providecommand \BibitemShut  [1]{\csname bibitem#1\endcsname}%
\let\auto@bib@innerbib\@empty
\bibitem [{\citenamefont {Abbott}\ \emph {et~al.}(2017{\natexlab{a}})\citenamefont {Abbott} \emph {et~al.}}]{LIGOScientific:2017vwq}%
  \BibitemOpen
  \bibfield  {author} {\bibinfo {author} {\bibfnamefont {B.~P.}\ \bibnamefont {Abbott}} \emph {et~al.} (\bibinfo {collaboration} {LIGO Scientific, Virgo}),\ }\href {\doibase 10.1103/PhysRevLett.119.161101} {\bibfield  {journal} {\bibinfo  {journal} {Phys. Rev. Lett.}\ }\textbf {\bibinfo {volume} {119}},\ \bibinfo {pages} {161101} (\bibinfo {year} {2017}{\natexlab{a}})},\ \Eprint {http://arxiv.org/abs/1710.05832} {arXiv:1710.05832 [gr-qc]} \BibitemShut {NoStop}%
\bibitem [{\citenamefont {Abbott}\ \emph {et~al.}(2018)\citenamefont {Abbott} \emph {et~al.}}]{PhysRevLett.121.161101}%
  \BibitemOpen
  \bibfield  {author} {\bibinfo {author} {\bibfnamefont {B.~P.}\ \bibnamefont {Abbott}} \emph {et~al.} (\bibinfo {collaboration} {The LIGO Scientific Collaboration and the Virgo Collaboration}),\ }\href {\doibase 10.1103/PhysRevLett.121.161101} {\bibfield  {journal} {\bibinfo  {journal} {Phys. Rev. Lett.}\ }\textbf {\bibinfo {volume} {121}},\ \bibinfo {pages} {161101} (\bibinfo {year} {2018})}\BibitemShut {NoStop}%
\bibitem [{\citenamefont {Miller}\ \emph {et~al.}(2019)\citenamefont {Miller} \emph {et~al.}}]{Miller:2019cac}%
  \BibitemOpen
  \bibfield  {author} {\bibinfo {author} {\bibfnamefont {M.~C.}\ \bibnamefont {Miller}} \emph {et~al.},\ }\href {\doibase 10.3847/2041-8213/ab50c5} {\bibfield  {journal} {\bibinfo  {journal} {Astrophys. J. Lett.}\ }\textbf {\bibinfo {volume} {887}},\ \bibinfo {pages} {L24} (\bibinfo {year} {2019})},\ \Eprint {http://arxiv.org/abs/1912.05705} {arXiv:1912.05705 [astro-ph.HE]} \BibitemShut {NoStop}%
\bibitem [{\citenamefont {Riley}\ \emph {et~al.}(2019)\citenamefont {Riley} \emph {et~al.}}]{Riley:2019yda}%
  \BibitemOpen
  \bibfield  {author} {\bibinfo {author} {\bibfnamefont {T.~E.}\ \bibnamefont {Riley}} \emph {et~al.},\ }\href {\doibase 10.3847/2041-8213/ab481c} {\bibfield  {journal} {\bibinfo  {journal} {Astrophys. J. Lett.}\ }\textbf {\bibinfo {volume} {887}},\ \bibinfo {pages} {L21} (\bibinfo {year} {2019})},\ \Eprint {http://arxiv.org/abs/1912.05702} {arXiv:1912.05702 [astro-ph.HE]} \BibitemShut {NoStop}%
\bibitem [{\citenamefont {Riley}\ \emph {et~al.}(2021)\citenamefont {Riley} \emph {et~al.}}]{Riley:2021pdl}%
  \BibitemOpen
  \bibfield  {author} {\bibinfo {author} {\bibfnamefont {T.~E.}\ \bibnamefont {Riley}} \emph {et~al.},\ }\href {\doibase 10.3847/2041-8213/ac0a81} {\bibfield  {journal} {\bibinfo  {journal} {Astrophys. J. Lett.}\ }\textbf {\bibinfo {volume} {918}},\ \bibinfo {pages} {L27} (\bibinfo {year} {2021})},\ \Eprint {http://arxiv.org/abs/2105.06980} {arXiv:2105.06980 [astro-ph.HE]} \BibitemShut {NoStop}%
\bibitem [{\citenamefont {Miller}\ \emph {et~al.}(2021)\citenamefont {Miller} \emph {et~al.}}]{Miller_2021}%
  \BibitemOpen
  \bibfield  {author} {\bibinfo {author} {\bibfnamefont {M.~C.}\ \bibnamefont {Miller}} \emph {et~al.},\ }\href {\doibase 10.3847/2041-8213/ac089b} {\bibfield  {journal} {\bibinfo  {journal} {The Astrophysical Journal Letters}\ }\textbf {\bibinfo {volume} {918}},\ \bibinfo {pages} {L28} (\bibinfo {year} {2021})}\BibitemShut {NoStop}%
\bibitem [{\citenamefont {Abbott}\ \emph {et~al.}(2017{\natexlab{b}})\citenamefont {Abbott} \emph {et~al.}}]{Abbott_2017}%
  \BibitemOpen
  \bibfield  {author} {\bibinfo {author} {\bibfnamefont {B.~P.}\ \bibnamefont {Abbott}} \emph {et~al.},\ }\href {\doibase 10.1088/1361-6382/aa51f4} {\bibfield  {journal} {\bibinfo  {journal} {Classical and Quantum Gravity}\ }\textbf {\bibinfo {volume} {34}},\ \bibinfo {pages} {044001} (\bibinfo {year} {2017}{\natexlab{b}})}\BibitemShut {NoStop}%
\bibitem [{\citenamefont {Maggiore}\ \emph {et~al.}(2020)\citenamefont {Maggiore}, \citenamefont {Broeck}, \citenamefont {Bartolo}, \citenamefont {Belgacem}, \citenamefont {Bertacca}, \citenamefont {Bizouard}, \citenamefont {Branchesi}, \citenamefont {Clesse}, \citenamefont {Foffa}, \citenamefont {García-Bellido}, \citenamefont {Grimm}, \citenamefont {Harms}, \citenamefont {Hinderer}, \citenamefont {Matarrese}, \citenamefont {Palomba}, \citenamefont {Peloso}, \citenamefont {Ricciardone},\ and\ \citenamefont {Sakellariadou}}]{Maggiore_2020}%
  \BibitemOpen
  \bibfield  {author} {\bibinfo {author} {\bibfnamefont {M.}~\bibnamefont {Maggiore}}, \bibinfo {author} {\bibfnamefont {C.~V.~D.}\ \bibnamefont {Broeck}}, \bibinfo {author} {\bibfnamefont {N.}~\bibnamefont {Bartolo}}, \bibinfo {author} {\bibfnamefont {E.}~\bibnamefont {Belgacem}}, \bibinfo {author} {\bibfnamefont {D.}~\bibnamefont {Bertacca}}, \bibinfo {author} {\bibfnamefont {M.~A.}\ \bibnamefont {Bizouard}}, \bibinfo {author} {\bibfnamefont {M.}~\bibnamefont {Branchesi}}, \bibinfo {author} {\bibfnamefont {S.}~\bibnamefont {Clesse}}, \bibinfo {author} {\bibfnamefont {S.}~\bibnamefont {Foffa}}, \bibinfo {author} {\bibfnamefont {J.}~\bibnamefont {García-Bellido}}, \bibinfo {author} {\bibfnamefont {S.}~\bibnamefont {Grimm}}, \bibinfo {author} {\bibfnamefont {J.}~\bibnamefont {Harms}}, \bibinfo {author} {\bibfnamefont {T.}~\bibnamefont {Hinderer}}, \bibinfo {author} {\bibfnamefont {S.}~\bibnamefont {Matarrese}}, \bibinfo {author} {\bibfnamefont {C.}~\bibnamefont {Palomba}}, \bibinfo {author} {\bibfnamefont
  {M.}~\bibnamefont {Peloso}}, \bibinfo {author} {\bibfnamefont {A.}~\bibnamefont {Ricciardone}}, \ and\ \bibinfo {author} {\bibfnamefont {M.}~\bibnamefont {Sakellariadou}},\ }\href {\doibase 10.1088/1475-7516/2020/03/050} {\bibfield  {journal} {\bibinfo  {journal} {Journal of Cosmology and Astroparticle Physics}\ }\textbf {\bibinfo {volume} {2020}},\ \bibinfo {pages} {050} (\bibinfo {year} {2020})}\BibitemShut {NoStop}%
\bibitem [{\citenamefont {Sekiguchi}\ \emph {et~al.}(2011{\natexlab{a}})\citenamefont {Sekiguchi}, \citenamefont {Kiuchi}, \citenamefont {Kyutoku},\ and\ \citenamefont {Shibata}}]{Sekiguchi:2011mc}%
  \BibitemOpen
  \bibfield  {author} {\bibinfo {author} {\bibfnamefont {Y.}~\bibnamefont {Sekiguchi}}, \bibinfo {author} {\bibfnamefont {K.}~\bibnamefont {Kiuchi}}, \bibinfo {author} {\bibfnamefont {K.}~\bibnamefont {Kyutoku}}, \ and\ \bibinfo {author} {\bibfnamefont {M.}~\bibnamefont {Shibata}},\ }\href {\doibase 10.1103/PhysRevLett.107.211101} {\bibfield  {journal} {\bibinfo  {journal} {Phys. Rev. Lett.}\ }\textbf {\bibinfo {volume} {107}},\ \bibinfo {pages} {211101} (\bibinfo {year} {2011}{\natexlab{a}})},\ \Eprint {http://arxiv.org/abs/1110.4442} {arXiv:1110.4442 [astro-ph.HE]} \BibitemShut {NoStop}%
\bibitem [{\citenamefont {Hotokezaka}\ \emph {et~al.}(2011)\citenamefont {Hotokezaka}, \citenamefont {Kyutoku}, \citenamefont {Okawa}, \citenamefont {Shibata},\ and\ \citenamefont {Kiuchi}}]{Hotokezaka:2011dh}%
  \BibitemOpen
  \bibfield  {author} {\bibinfo {author} {\bibfnamefont {K.}~\bibnamefont {Hotokezaka}}, \bibinfo {author} {\bibfnamefont {K.}~\bibnamefont {Kyutoku}}, \bibinfo {author} {\bibfnamefont {H.}~\bibnamefont {Okawa}}, \bibinfo {author} {\bibfnamefont {M.}~\bibnamefont {Shibata}}, \ and\ \bibinfo {author} {\bibfnamefont {K.}~\bibnamefont {Kiuchi}},\ }\href {\doibase 10.1103/PhysRevD.83.124008} {\bibfield  {journal} {\bibinfo  {journal} {Phys. Rev. D}\ }\textbf {\bibinfo {volume} {83}},\ \bibinfo {pages} {124008} (\bibinfo {year} {2011})},\ \Eprint {http://arxiv.org/abs/1105.4370} {arXiv:1105.4370 [astro-ph.HE]} \BibitemShut {NoStop}%
\bibitem [{\citenamefont {Bauswein}\ and\ \citenamefont {Janka}(2012)}]{Bauswein:2011tp}%
  \BibitemOpen
  \bibfield  {author} {\bibinfo {author} {\bibfnamefont {A.}~\bibnamefont {Bauswein}}\ and\ \bibinfo {author} {\bibfnamefont {H.~T.}\ \bibnamefont {Janka}},\ }\href {\doibase 10.1103/PhysRevLett.108.011101} {\bibfield  {journal} {\bibinfo  {journal} {Phys. Rev. Lett.}\ }\textbf {\bibinfo {volume} {108}},\ \bibinfo {pages} {011101} (\bibinfo {year} {2012})},\ \Eprint {http://arxiv.org/abs/1106.1616} {arXiv:1106.1616 [astro-ph.SR]} \BibitemShut {NoStop}%
\bibitem [{\citenamefont {Takami}\ \emph {et~al.}(2014)\citenamefont {Takami}, \citenamefont {Rezzolla},\ and\ \citenamefont {Baiotti}}]{Takami:2014zpa}%
  \BibitemOpen
  \bibfield  {author} {\bibinfo {author} {\bibfnamefont {K.}~\bibnamefont {Takami}}, \bibinfo {author} {\bibfnamefont {L.}~\bibnamefont {Rezzolla}}, \ and\ \bibinfo {author} {\bibfnamefont {L.}~\bibnamefont {Baiotti}},\ }\href {\doibase 10.1103/PhysRevLett.113.091104} {\bibfield  {journal} {\bibinfo  {journal} {Phys. Rev. Lett.}\ }\textbf {\bibinfo {volume} {113}},\ \bibinfo {pages} {091104} (\bibinfo {year} {2014})},\ \Eprint {http://arxiv.org/abs/1403.5672} {arXiv:1403.5672 [gr-qc]} \BibitemShut {NoStop}%
\bibitem [{\citenamefont {Bernuzzi}\ \emph {et~al.}(2015)\citenamefont {Bernuzzi}, \citenamefont {Dietrich},\ and\ \citenamefont {Nagar}}]{Bernuzzi2015}%
  \BibitemOpen
  \bibfield  {author} {\bibinfo {author} {\bibfnamefont {S.}~\bibnamefont {Bernuzzi}}, \bibinfo {author} {\bibfnamefont {T.}~\bibnamefont {Dietrich}}, \ and\ \bibinfo {author} {\bibfnamefont {A.}~\bibnamefont {Nagar}},\ }\href {\doibase 10.1103/PhysRevLett.115.091101} {\bibfield  {journal} {\bibinfo  {journal} {Phys. Rev. Lett.}\ }\textbf {\bibinfo {volume} {115}},\ \bibinfo {pages} {091101} (\bibinfo {year} {2015})}\BibitemShut {NoStop}%
\bibitem [{\citenamefont {Bauswein}\ and\ \citenamefont {Stergioulas}(2015)}]{Bauswein:2015yca}%
  \BibitemOpen
  \bibfield  {author} {\bibinfo {author} {\bibfnamefont {A.}~\bibnamefont {Bauswein}}\ and\ \bibinfo {author} {\bibfnamefont {N.}~\bibnamefont {Stergioulas}},\ }\href {\doibase 10.1103/PhysRevD.91.124056} {\bibfield  {journal} {\bibinfo  {journal} {Phys. Rev. D}\ }\textbf {\bibinfo {volume} {91}},\ \bibinfo {pages} {124056} (\bibinfo {year} {2015})},\ \Eprint {http://arxiv.org/abs/1502.03176} {arXiv:1502.03176 [astro-ph.SR]} \BibitemShut {NoStop}%
\bibitem [{\citenamefont {Rezzolla}\ and\ \citenamefont {Takami}(2016)}]{Rezzolla:2016nxn}%
  \BibitemOpen
  \bibfield  {author} {\bibinfo {author} {\bibfnamefont {L.}~\bibnamefont {Rezzolla}}\ and\ \bibinfo {author} {\bibfnamefont {K.}~\bibnamefont {Takami}},\ }\href {\doibase 10.1103/PhysRevD.93.124051} {\bibfield  {journal} {\bibinfo  {journal} {Phys. Rev. D}\ }\textbf {\bibinfo {volume} {93}},\ \bibinfo {pages} {124051} (\bibinfo {year} {2016})},\ \Eprint {http://arxiv.org/abs/1604.00246} {arXiv:1604.00246 [gr-qc]} \BibitemShut {NoStop}%
\bibitem [{\citenamefont {Maione}\ \emph {et~al.}(2016)\citenamefont {Maione}, \citenamefont {Pietri}, \citenamefont {Feo},\ and\ \citenamefont {Löffler}}]{Maione_2016}%
  \BibitemOpen
  \bibfield  {author} {\bibinfo {author} {\bibfnamefont {F.}~\bibnamefont {Maione}}, \bibinfo {author} {\bibfnamefont {R.~D.}\ \bibnamefont {Pietri}}, \bibinfo {author} {\bibfnamefont {A.}~\bibnamefont {Feo}}, \ and\ \bibinfo {author} {\bibfnamefont {F.}~\bibnamefont {Löffler}},\ }\href {\doibase 10.1088/0264-9381/33/17/175009} {\bibfield  {journal} {\bibinfo  {journal} {Classical and Quantum Gravity}\ }\textbf {\bibinfo {volume} {33}},\ \bibinfo {pages} {175009} (\bibinfo {year} {2016})}\BibitemShut {NoStop}%
\bibitem [{\citenamefont {Vretinaris}\ \emph {et~al.}(2020)\citenamefont {Vretinaris}, \citenamefont {Stergioulas},\ and\ \citenamefont {Bauswein}}]{Vretinaris:2019spn}%
  \BibitemOpen
  \bibfield  {author} {\bibinfo {author} {\bibfnamefont {S.}~\bibnamefont {Vretinaris}}, \bibinfo {author} {\bibfnamefont {N.}~\bibnamefont {Stergioulas}}, \ and\ \bibinfo {author} {\bibfnamefont {A.}~\bibnamefont {Bauswein}},\ }\href {\doibase 10.1103/PhysRevD.101.084039} {\bibfield  {journal} {\bibinfo  {journal} {Phys. Rev. D}\ }\textbf {\bibinfo {volume} {101}},\ \bibinfo {pages} {084039} (\bibinfo {year} {2020})},\ \Eprint {http://arxiv.org/abs/1910.10856} {arXiv:1910.10856 [gr-qc]} \BibitemShut {NoStop}%
\bibitem [{\citenamefont {Soultanis}\ \emph {et~al.}(2022)\citenamefont {Soultanis}, \citenamefont {Bauswein},\ and\ \citenamefont {Stergioulas}}]{Soultanis:2021oia}%
  \BibitemOpen
  \bibfield  {author} {\bibinfo {author} {\bibfnamefont {T.}~\bibnamefont {Soultanis}}, \bibinfo {author} {\bibfnamefont {A.}~\bibnamefont {Bauswein}}, \ and\ \bibinfo {author} {\bibfnamefont {N.}~\bibnamefont {Stergioulas}},\ }\href {\doibase 10.1103/PhysRevD.105.043020} {\bibfield  {journal} {\bibinfo  {journal} {Phys. Rev. D}\ }\textbf {\bibinfo {volume} {105}},\ \bibinfo {pages} {043020} (\bibinfo {year} {2022})},\ \Eprint {http://arxiv.org/abs/2111.08353} {arXiv:2111.08353 [astro-ph.HE]} \BibitemShut {NoStop}%
\bibitem [{\citenamefont {Stergioulas}\ \emph {et~al.}(2011)\citenamefont {Stergioulas}, \citenamefont {Bauswein}, \citenamefont {Zagkouris},\ and\ \citenamefont {Janka}}]{Stergioulas:2011gd}%
  \BibitemOpen
  \bibfield  {author} {\bibinfo {author} {\bibfnamefont {N.}~\bibnamefont {Stergioulas}}, \bibinfo {author} {\bibfnamefont {A.}~\bibnamefont {Bauswein}}, \bibinfo {author} {\bibfnamefont {K.}~\bibnamefont {Zagkouris}}, \ and\ \bibinfo {author} {\bibfnamefont {H.-T.}\ \bibnamefont {Janka}},\ }\href {\doibase 10.1111/j.1365-2966.2011.19493.x} {\bibfield  {journal} {\bibinfo  {journal} {Mon. Not. Roy. Astron. Soc.}\ }\textbf {\bibinfo {volume} {418}},\ \bibinfo {pages} {427} (\bibinfo {year} {2011})},\ \Eprint {http://arxiv.org/abs/1105.0368} {arXiv:1105.0368 [gr-qc]} \BibitemShut {NoStop}%
\bibitem [{\citenamefont {Bauswein}\ \emph {et~al.}(2019)\citenamefont {Bauswein}, \citenamefont {Bastian}, \citenamefont {Blaschke}, \citenamefont {Chatziioannou}, \citenamefont {Clark}, \citenamefont {Fischer},\ and\ \citenamefont {Oertel}}]{Bauswein:2018bma}%
  \BibitemOpen
  \bibfield  {author} {\bibinfo {author} {\bibfnamefont {A.}~\bibnamefont {Bauswein}}, \bibinfo {author} {\bibfnamefont {N.-U.~F.}\ \bibnamefont {Bastian}}, \bibinfo {author} {\bibfnamefont {D.~B.}\ \bibnamefont {Blaschke}}, \bibinfo {author} {\bibfnamefont {K.}~\bibnamefont {Chatziioannou}}, \bibinfo {author} {\bibfnamefont {J.~A.}\ \bibnamefont {Clark}}, \bibinfo {author} {\bibfnamefont {T.}~\bibnamefont {Fischer}}, \ and\ \bibinfo {author} {\bibfnamefont {M.}~\bibnamefont {Oertel}},\ }\href {\doibase 10.1103/PhysRevLett.122.061102} {\bibfield  {journal} {\bibinfo  {journal} {Phys. Rev. Lett.}\ }\textbf {\bibinfo {volume} {122}},\ \bibinfo {pages} {061102} (\bibinfo {year} {2019})},\ \Eprint {http://arxiv.org/abs/1809.01116} {arXiv:1809.01116 [astro-ph.HE]} \BibitemShut {NoStop}%
\bibitem [{\citenamefont {Blacker}\ \emph {et~al.}(2020)\citenamefont {Blacker}, \citenamefont {Bastian}, \citenamefont {Bauswein}, \citenamefont {Blaschke}, \citenamefont {Fischer}, \citenamefont {Oertel}, \citenamefont {Soultanis},\ and\ \citenamefont {Typel}}]{Blacker:2020nlq}%
  \BibitemOpen
  \bibfield  {author} {\bibinfo {author} {\bibfnamefont {S.}~\bibnamefont {Blacker}}, \bibinfo {author} {\bibfnamefont {N.-U.~F.}\ \bibnamefont {Bastian}}, \bibinfo {author} {\bibfnamefont {A.}~\bibnamefont {Bauswein}}, \bibinfo {author} {\bibfnamefont {D.~B.}\ \bibnamefont {Blaschke}}, \bibinfo {author} {\bibfnamefont {T.}~\bibnamefont {Fischer}}, \bibinfo {author} {\bibfnamefont {M.}~\bibnamefont {Oertel}}, \bibinfo {author} {\bibfnamefont {T.}~\bibnamefont {Soultanis}}, \ and\ \bibinfo {author} {\bibfnamefont {S.}~\bibnamefont {Typel}},\ }\href {\doibase 10.1103/PhysRevD.102.123023} {\bibfield  {journal} {\bibinfo  {journal} {Phys. Rev. D}\ }\textbf {\bibinfo {volume} {102}},\ \bibinfo {pages} {123023} (\bibinfo {year} {2020})},\ \Eprint {http://arxiv.org/abs/2006.03789} {arXiv:2006.03789 [astro-ph.HE]} \BibitemShut {NoStop}%
\bibitem [{\citenamefont {Weih}\ \emph {et~al.}(2020)\citenamefont {Weih}, \citenamefont {Hanauske},\ and\ \citenamefont {Rezzolla}}]{Weih:2019xvw}%
  \BibitemOpen
  \bibfield  {author} {\bibinfo {author} {\bibfnamefont {L.~R.}\ \bibnamefont {Weih}}, \bibinfo {author} {\bibfnamefont {M.}~\bibnamefont {Hanauske}}, \ and\ \bibinfo {author} {\bibfnamefont {L.}~\bibnamefont {Rezzolla}},\ }\href {\doibase 10.1103/PhysRevLett.124.171103} {\bibfield  {journal} {\bibinfo  {journal} {Phys. Rev. Lett.}\ }\textbf {\bibinfo {volume} {124}},\ \bibinfo {pages} {171103} (\bibinfo {year} {2020})},\ \Eprint {http://arxiv.org/abs/1912.09340} {arXiv:1912.09340 [gr-qc]} \BibitemShut {NoStop}%
\bibitem [{\citenamefont {Huang}\ \emph {et~al.}(2022)\citenamefont {Huang}, \citenamefont {Baiotti}, \citenamefont {Kojo}, \citenamefont {Takami}, \citenamefont {Sotani}, \citenamefont {Togashi}, \citenamefont {Hatsuda}, \citenamefont {Nagataki},\ and\ \citenamefont {Fan}}]{Huang:2022mqp}%
  \BibitemOpen
  \bibfield  {author} {\bibinfo {author} {\bibfnamefont {Y.-J.}\ \bibnamefont {Huang}}, \bibinfo {author} {\bibfnamefont {L.}~\bibnamefont {Baiotti}}, \bibinfo {author} {\bibfnamefont {T.}~\bibnamefont {Kojo}}, \bibinfo {author} {\bibfnamefont {K.}~\bibnamefont {Takami}}, \bibinfo {author} {\bibfnamefont {H.}~\bibnamefont {Sotani}}, \bibinfo {author} {\bibfnamefont {H.}~\bibnamefont {Togashi}}, \bibinfo {author} {\bibfnamefont {T.}~\bibnamefont {Hatsuda}}, \bibinfo {author} {\bibfnamefont {S.}~\bibnamefont {Nagataki}}, \ and\ \bibinfo {author} {\bibfnamefont {Y.-Z.}\ \bibnamefont {Fan}},\ }\href {\doibase 10.1103/PhysRevLett.129.181101} {\bibfield  {journal} {\bibinfo  {journal} {Phys. Rev. Lett.}\ }\textbf {\bibinfo {volume} {129}},\ \bibinfo {pages} {181101} (\bibinfo {year} {2022})},\ \Eprint {http://arxiv.org/abs/2203.04528} {arXiv:2203.04528 [astro-ph.HE]} \BibitemShut {NoStop}%
\bibitem [{\citenamefont {Masuda}\ \emph {et~al.}(2013)\citenamefont {Masuda}, \citenamefont {Hatsuda},\ and\ \citenamefont {Takatsuka}}]{Masuda2013}%
  \BibitemOpen
  \bibfield  {author} {\bibinfo {author} {\bibfnamefont {K.}~\bibnamefont {Masuda}}, \bibinfo {author} {\bibfnamefont {T.}~\bibnamefont {Hatsuda}}, \ and\ \bibinfo {author} {\bibfnamefont {T.}~\bibnamefont {Takatsuka}},\ }\href {\doibase 10.1088/0004-637X/764/1/12} {\bibfield  {journal} {\bibinfo  {journal} {The Astrophysical Journal}\ }\textbf {\bibinfo {volume} {764}},\ \bibinfo {pages} {12} (\bibinfo {year} {2013})}\BibitemShut {NoStop}%
\bibitem [{\citenamefont {Baym}\ \emph {et~al.}(2018)\citenamefont {Baym}, \citenamefont {Hatsuda}, \citenamefont {Kojo}, \citenamefont {Powell}, \citenamefont {Song},\ and\ \citenamefont {Takatsuka}}]{Baym2018}%
  \BibitemOpen
  \bibfield  {author} {\bibinfo {author} {\bibfnamefont {G.}~\bibnamefont {Baym}}, \bibinfo {author} {\bibfnamefont {T.}~\bibnamefont {Hatsuda}}, \bibinfo {author} {\bibfnamefont {T.}~\bibnamefont {Kojo}}, \bibinfo {author} {\bibfnamefont {P.~D.}\ \bibnamefont {Powell}}, \bibinfo {author} {\bibfnamefont {Y.}~\bibnamefont {Song}}, \ and\ \bibinfo {author} {\bibfnamefont {T.}~\bibnamefont {Takatsuka}},\ }\href {\doibase 10.1088/1361-6633/aaae14} {\bibfield  {journal} {\bibinfo  {journal} {Reports on Progress in Physics}\ }\textbf {\bibinfo {volume} {81}},\ \bibinfo {pages} {056902} (\bibinfo {year} {2018})}\BibitemShut {NoStop}%
\bibitem [{\citenamefont {Baym}\ \emph {et~al.}(2019)\citenamefont {Baym}, \citenamefont {Furusawa}, \citenamefont {Hatsuda}, \citenamefont {Kojo},\ and\ \citenamefont {Togashi}}]{Baym:2019iky}%
  \BibitemOpen
  \bibfield  {author} {\bibinfo {author} {\bibfnamefont {G.}~\bibnamefont {Baym}}, \bibinfo {author} {\bibfnamefont {S.}~\bibnamefont {Furusawa}}, \bibinfo {author} {\bibfnamefont {T.}~\bibnamefont {Hatsuda}}, \bibinfo {author} {\bibfnamefont {T.}~\bibnamefont {Kojo}}, \ and\ \bibinfo {author} {\bibfnamefont {H.}~\bibnamefont {Togashi}},\ }\href {\doibase 10.3847/1538-4357/ab441e} {\bibfield  {journal} {\bibinfo  {journal} {Astrophys. J.}\ }\textbf {\bibinfo {volume} {885}},\ \bibinfo {pages} {42} (\bibinfo {year} {2019})},\ \Eprint {http://arxiv.org/abs/1903.08963} {arXiv:1903.08963 [astro-ph.HE]} \BibitemShut {NoStop}%
\bibitem [{\citenamefont {Kojo}\ \emph {et~al.}(2022)\citenamefont {Kojo}, \citenamefont {Baym},\ and\ \citenamefont {Hatsuda}}]{Kojo2022}%
  \BibitemOpen
  \bibfield  {author} {\bibinfo {author} {\bibfnamefont {T.}~\bibnamefont {Kojo}}, \bibinfo {author} {\bibfnamefont {G.}~\bibnamefont {Baym}}, \ and\ \bibinfo {author} {\bibfnamefont {T.}~\bibnamefont {Hatsuda}},\ }\href {\doibase 10.3847/1538-4357/ac7876} {\bibfield  {journal} {\bibinfo  {journal} {The Astrophysical Journal}\ }\textbf {\bibinfo {volume} {934}},\ \bibinfo {pages} {46} (\bibinfo {year} {2022})}\BibitemShut {NoStop}%
\bibitem [{\citenamefont {Yao}\ \emph {et~al.}(2024)\citenamefont {Yao}, \citenamefont {Sorensen}, \citenamefont {Dexheimer},\ and\ \citenamefont {Noronha-Hostler}}]{PhysRevC.109.065803}%
  \BibitemOpen
  \bibfield  {author} {\bibinfo {author} {\bibfnamefont {N.}~\bibnamefont {Yao}}, \bibinfo {author} {\bibfnamefont {A.}~\bibnamefont {Sorensen}}, \bibinfo {author} {\bibfnamefont {V.}~\bibnamefont {Dexheimer}}, \ and\ \bibinfo {author} {\bibfnamefont {J.}~\bibnamefont {Noronha-Hostler}},\ }\href {\doibase 10.1103/PhysRevC.109.065803} {\bibfield  {journal} {\bibinfo  {journal} {Phys. Rev. C}\ }\textbf {\bibinfo {volume} {109}},\ \bibinfo {pages} {065803} (\bibinfo {year} {2024})}\BibitemShut {NoStop}%
\bibitem [{\citenamefont {Kiuchi}\ \emph {et~al.}(2014)\citenamefont {Kiuchi}, \citenamefont {Kyutoku}, \citenamefont {Sekiguchi}, \citenamefont {Shibata},\ and\ \citenamefont {Wada}}]{Kiuchi:2014hja}%
  \BibitemOpen
  \bibfield  {author} {\bibinfo {author} {\bibfnamefont {K.}~\bibnamefont {Kiuchi}}, \bibinfo {author} {\bibfnamefont {K.}~\bibnamefont {Kyutoku}}, \bibinfo {author} {\bibfnamefont {Y.}~\bibnamefont {Sekiguchi}}, \bibinfo {author} {\bibfnamefont {M.}~\bibnamefont {Shibata}}, \ and\ \bibinfo {author} {\bibfnamefont {T.}~\bibnamefont {Wada}},\ }\href {\doibase 10.1103/PhysRevD.90.041502} {\bibfield  {journal} {\bibinfo  {journal} {Phys. Rev. D}\ }\textbf {\bibinfo {volume} {90}},\ \bibinfo {pages} {041502(R)} (\bibinfo {year} {2014})},\ \Eprint {http://arxiv.org/abs/1407.2660} {arXiv:1407.2660 [astro-ph.HE]} \BibitemShut {NoStop}%
\bibitem [{\citenamefont {Aguilera-Miret}\ \emph {et~al.}(2023)\citenamefont {Aguilera-Miret}, \citenamefont {Palenzuela}, \citenamefont {Carrasco},\ and\ \citenamefont {Vigan\`o}}]{Aguilera-Miret:2023qih}%
  \BibitemOpen
  \bibfield  {author} {\bibinfo {author} {\bibfnamefont {R.}~\bibnamefont {Aguilera-Miret}}, \bibinfo {author} {\bibfnamefont {C.}~\bibnamefont {Palenzuela}}, \bibinfo {author} {\bibfnamefont {F.}~\bibnamefont {Carrasco}}, \ and\ \bibinfo {author} {\bibfnamefont {D.}~\bibnamefont {Vigan\`o}},\ }\href {\doibase 10.1103/PhysRevD.108.103001} {\bibfield  {journal} {\bibinfo  {journal} {Phys. Rev. D}\ }\textbf {\bibinfo {volume} {108}},\ \bibinfo {pages} {103001} (\bibinfo {year} {2023})},\ \Eprint {http://arxiv.org/abs/2307.04837} {arXiv:2307.04837 [astro-ph.HE]} \BibitemShut {NoStop}%
\bibitem [{\citenamefont {Douchin}\ and\ \citenamefont {Haensel}(2001)}]{Douchin:2001sv}%
  \BibitemOpen
  \bibfield  {author} {\bibinfo {author} {\bibfnamefont {F.}~\bibnamefont {Douchin}}\ and\ \bibinfo {author} {\bibfnamefont {P.}~\bibnamefont {Haensel}},\ }\href {\doibase 10.1051/0004-6361:20011402} {\bibfield  {journal} {\bibinfo  {journal} {Astron. Astrophys.}\ }\textbf {\bibinfo {volume} {380}},\ \bibinfo {pages} {151} (\bibinfo {year} {2001})},\ \Eprint {http://arxiv.org/abs/astro-ph/0111092} {arXiv:astro-ph/0111092} \BibitemShut {NoStop}%
\bibitem [{\citenamefont {Alford}\ \emph {et~al.}(2005)\citenamefont {Alford}, \citenamefont {Braby}, \citenamefont {Paris},\ and\ \citenamefont {Reddy}}]{Alford:2004pf}%
  \BibitemOpen
  \bibfield  {author} {\bibinfo {author} {\bibfnamefont {M.}~\bibnamefont {Alford}}, \bibinfo {author} {\bibfnamefont {M.}~\bibnamefont {Braby}}, \bibinfo {author} {\bibfnamefont {M.~W.}\ \bibnamefont {Paris}}, \ and\ \bibinfo {author} {\bibfnamefont {S.}~\bibnamefont {Reddy}},\ }\href {\doibase 10.1086/430902} {\bibfield  {journal} {\bibinfo  {journal} {Astrophys. J.}\ }\textbf {\bibinfo {volume} {629}},\ \bibinfo {pages} {969} (\bibinfo {year} {2005})},\ \Eprint {http://arxiv.org/abs/nucl-th/0411016} {arXiv:nucl-th/0411016} \BibitemShut {NoStop}%
\bibitem [{\citenamefont {{Cook}}\ \emph {et~al.}(1992)\citenamefont {{Cook}}, \citenamefont {{Shapiro}},\ and\ \citenamefont {{Teukolsky}}}]{Cook92b}%
  \BibitemOpen
  \bibfield  {author} {\bibinfo {author} {\bibfnamefont {G.~B.}\ \bibnamefont {{Cook}}}, \bibinfo {author} {\bibfnamefont {S.~L.}\ \bibnamefont {{Shapiro}}}, \ and\ \bibinfo {author} {\bibfnamefont {S.~A.}\ \bibnamefont {{Teukolsky}}},\ }\href@noop {} {\bibfield  {journal} {\bibinfo  {journal} {Astrophys. J.}\ }\textbf {\bibinfo {volume} {398}},\ \bibinfo {pages} {203} (\bibinfo {year} {1992})}\BibitemShut {NoStop}%
\bibitem [{\citenamefont {Baumgarte}\ \emph {et~al.}(2000)\citenamefont {Baumgarte}, \citenamefont {Shapiro},\ and\ \citenamefont {Shibata}}]{Baumgarte:1999cq}%
  \BibitemOpen
  \bibfield  {author} {\bibinfo {author} {\bibfnamefont {T.~W.}\ \bibnamefont {Baumgarte}}, \bibinfo {author} {\bibfnamefont {S.~L.}\ \bibnamefont {Shapiro}}, \ and\ \bibinfo {author} {\bibfnamefont {M.}~\bibnamefont {Shibata}},\ }\href {\doibase 10.1086/312425} {\bibfield  {journal} {\bibinfo  {journal} {Astrophys. J. Lett.}\ }\textbf {\bibinfo {volume} {528}},\ \bibinfo {pages} {L29} (\bibinfo {year} {2000})},\ \Eprint {http://arxiv.org/abs/astro-ph/9910565} {arXiv:astro-ph/9910565} \BibitemShut {NoStop}%
\bibitem [{\citenamefont {Siegel}\ \emph {et~al.}(2013)\citenamefont {Siegel}, \citenamefont {Ciolfi}, \citenamefont {Harte},\ and\ \citenamefont {Rezzolla}}]{Siegel:2013nrw}%
  \BibitemOpen
  \bibfield  {author} {\bibinfo {author} {\bibfnamefont {D.~M.}\ \bibnamefont {Siegel}}, \bibinfo {author} {\bibfnamefont {R.}~\bibnamefont {Ciolfi}}, \bibinfo {author} {\bibfnamefont {A.~I.}\ \bibnamefont {Harte}}, \ and\ \bibinfo {author} {\bibfnamefont {L.}~\bibnamefont {Rezzolla}},\ }\href {\doibase 10.1103/PhysRevD.87.121302} {\bibfield  {journal} {\bibinfo  {journal} {Phys. Rev. D}\ }\textbf {\bibinfo {volume} {87}},\ \bibinfo {pages} {121302(R)} (\bibinfo {year} {2013})},\ \Eprint {http://arxiv.org/abs/1302.4368} {arXiv:1302.4368 [gr-qc]} \BibitemShut {NoStop}%
\bibitem [{\citenamefont {Ciolfi}\ \emph {et~al.}(2019)\citenamefont {Ciolfi}, \citenamefont {Kastaun}, \citenamefont {Kalinani},\ and\ \citenamefont {Giacomazzo}}]{Ciolfi:2019fie}%
  \BibitemOpen
  \bibfield  {author} {\bibinfo {author} {\bibfnamefont {R.}~\bibnamefont {Ciolfi}}, \bibinfo {author} {\bibfnamefont {W.}~\bibnamefont {Kastaun}}, \bibinfo {author} {\bibfnamefont {J.~V.}\ \bibnamefont {Kalinani}}, \ and\ \bibinfo {author} {\bibfnamefont {B.}~\bibnamefont {Giacomazzo}},\ }\href {\doibase 10.1103/PhysRevD.100.023005} {\bibfield  {journal} {\bibinfo  {journal} {Phys. Rev. D}\ }\textbf {\bibinfo {volume} {100}},\ \bibinfo {pages} {023005} (\bibinfo {year} {2019})},\ \Eprint {http://arxiv.org/abs/1904.10222} {arXiv:1904.10222 [astro-ph.HE]} \BibitemShut {NoStop}%
\bibitem [{\citenamefont {Ruiz}\ \emph {et~al.}(2020)\citenamefont {Ruiz}, \citenamefont {Tsokaros},\ and\ \citenamefont {Shapiro}}]{Ruiz:2020via}%
  \BibitemOpen
  \bibfield  {author} {\bibinfo {author} {\bibfnamefont {M.}~\bibnamefont {Ruiz}}, \bibinfo {author} {\bibfnamefont {A.}~\bibnamefont {Tsokaros}}, \ and\ \bibinfo {author} {\bibfnamefont {S.~L.}\ \bibnamefont {Shapiro}},\ }\href {\doibase 10.1103/PhysRevD.101.064042} {\bibfield  {journal} {\bibinfo  {journal} {Phys. Rev. D}\ }\textbf {\bibinfo {volume} {101}},\ \bibinfo {pages} {064042} (\bibinfo {year} {2020})},\ \Eprint {http://arxiv.org/abs/2001.09153} {arXiv:2001.09153 [astro-ph.HE]} \BibitemShut {NoStop}%
\bibitem [{\citenamefont {Ruiz}\ \emph {et~al.}(2021)\citenamefont {Ruiz}, \citenamefont {Tsokaros},\ and\ \citenamefont {Shapiro}}]{Ruiz:2021qmm}%
  \BibitemOpen
  \bibfield  {author} {\bibinfo {author} {\bibfnamefont {M.}~\bibnamefont {Ruiz}}, \bibinfo {author} {\bibfnamefont {A.}~\bibnamefont {Tsokaros}}, \ and\ \bibinfo {author} {\bibfnamefont {S.~L.}\ \bibnamefont {Shapiro}},\ }\href {\doibase 10.1103/PhysRevD.104.124049} {\bibfield  {journal} {\bibinfo  {journal} {Phys. Rev. D}\ }\textbf {\bibinfo {volume} {104}},\ \bibinfo {pages} {124049} (\bibinfo {year} {2021})},\ \Eprint {http://arxiv.org/abs/2110.11968} {arXiv:2110.11968 [astro-ph.HE]} \BibitemShut {NoStop}%
\bibitem [{\citenamefont {Bamber}\ \emph {et~al.}(2024{\natexlab{a}})\citenamefont {Bamber}, \citenamefont {Tsokaros}, \citenamefont {Ruiz},\ and\ \citenamefont {Shapiro}}]{Bamber:2024kfb}%
  \BibitemOpen
  \bibfield  {author} {\bibinfo {author} {\bibfnamefont {J.}~\bibnamefont {Bamber}}, \bibinfo {author} {\bibfnamefont {A.}~\bibnamefont {Tsokaros}}, \bibinfo {author} {\bibfnamefont {M.}~\bibnamefont {Ruiz}}, \ and\ \bibinfo {author} {\bibfnamefont {S.~L.}\ \bibnamefont {Shapiro}},\ }\href {\doibase 10.1103/PhysRevD.110.024046} {\bibfield  {journal} {\bibinfo  {journal} {Phys. Rev. D}\ }\textbf {\bibinfo {volume} {110}},\ \bibinfo {pages} {024046} (\bibinfo {year} {2024}{\natexlab{a}})},\ \Eprint {http://arxiv.org/abs/2405.03705} {arXiv:2405.03705 [astro-ph.HE]} \BibitemShut {NoStop}%
\bibitem [{\citenamefont {Kiuchi}\ \emph {et~al.}(2024)\citenamefont {Kiuchi}, \citenamefont {Reboul-Salze}, \citenamefont {Shibata},\ and\ \citenamefont {Sekiguchi}}]{Kiuchi:2023obe}%
  \BibitemOpen
  \bibfield  {author} {\bibinfo {author} {\bibfnamefont {K.}~\bibnamefont {Kiuchi}}, \bibinfo {author} {\bibfnamefont {A.}~\bibnamefont {Reboul-Salze}}, \bibinfo {author} {\bibfnamefont {M.}~\bibnamefont {Shibata}}, \ and\ \bibinfo {author} {\bibfnamefont {Y.}~\bibnamefont {Sekiguchi}},\ }\href {\doibase 10.1038/s41550-024-02194-y} {\bibfield  {journal} {\bibinfo  {journal} {Nature Astron.}\ }\textbf {\bibinfo {volume} {8}},\ \bibinfo {pages} {298} (\bibinfo {year} {2024})},\ \Eprint {http://arxiv.org/abs/2306.15721} {arXiv:2306.15721 [astro-ph.HE]} \BibitemShut {NoStop}%
\bibitem [{\citenamefont {Bamber}\ \emph {et~al.}(2024{\natexlab{b}})\citenamefont {Bamber}, \citenamefont {Tsokaros}, \citenamefont {Ruiz},\ and\ \citenamefont {Shapiro}}]{Bamber2024un}%
  \BibitemOpen
  \bibfield  {author} {\bibinfo {author} {\bibfnamefont {J.}~\bibnamefont {Bamber}}, \bibinfo {author} {\bibfnamefont {A.}~\bibnamefont {Tsokaros}}, \bibinfo {author} {\bibfnamefont {M.}~\bibnamefont {Ruiz}}, \ and\ \bibinfo {author} {\bibfnamefont {S.~L.}\ \bibnamefont {Shapiro}},\ }\href@noop {} {\enquote {\bibinfo {title} {{Post-merger gravitational wave signals from binary neutron stars: Effect of the magnetic field}},}\ } (\bibinfo {year} {2024}{\natexlab{b}}),\ \bibinfo {note} {submitted for publication}\BibitemShut {NoStop}%
\bibitem [{\citenamefont {Price}\ and\ \citenamefont {Rosswog}(2006)}]{Price:2006fi}%
  \BibitemOpen
  \bibfield  {author} {\bibinfo {author} {\bibfnamefont {D.}~\bibnamefont {Price}}\ and\ \bibinfo {author} {\bibfnamefont {S.}~\bibnamefont {Rosswog}},\ }\href {\doibase 10.1126/science.1125201} {\bibfield  {journal} {\bibinfo  {journal} {Science}\ }\textbf {\bibinfo {volume} {312}},\ \bibinfo {pages} {719} (\bibinfo {year} {2006})},\ \Eprint {http://arxiv.org/abs/astro-ph/0603845} {arXiv:astro-ph/0603845} \BibitemShut {NoStop}%
\bibitem [{\citenamefont {Kiuchi}\ \emph {et~al.}(2015)\citenamefont {Kiuchi}, \citenamefont {Cerd\'a-Dur\'an}, \citenamefont {Kyutoku}, \citenamefont {Sekiguchi},\ and\ \citenamefont {Shibata}}]{Kiuchi:2015sga}%
  \BibitemOpen
  \bibfield  {author} {\bibinfo {author} {\bibfnamefont {K.}~\bibnamefont {Kiuchi}}, \bibinfo {author} {\bibfnamefont {P.}~\bibnamefont {Cerd\'a-Dur\'an}}, \bibinfo {author} {\bibfnamefont {K.}~\bibnamefont {Kyutoku}}, \bibinfo {author} {\bibfnamefont {Y.}~\bibnamefont {Sekiguchi}}, \ and\ \bibinfo {author} {\bibfnamefont {M.}~\bibnamefont {Shibata}},\ }\href {\doibase 10.1103/PhysRevD.92.124034} {\bibfield  {journal} {\bibinfo  {journal} {Phys. Rev. D}\ }\textbf {\bibinfo {volume} {92}},\ \bibinfo {pages} {124034} (\bibinfo {year} {2015})},\ \Eprint {http://arxiv.org/abs/1509.09205} {arXiv:1509.09205 [astro-ph.HE]} \BibitemShut {NoStop}%
\bibitem [{\citenamefont {Aguilera-Miret}\ \emph {et~al.}(2020)\citenamefont {Aguilera-Miret}, \citenamefont {Vigan\`o}, \citenamefont {Carrasco}, \citenamefont {Mi\~nano},\ and\ \citenamefont {Palenzuela}}]{Aguilera-Miret:2020dhz}%
  \BibitemOpen
  \bibfield  {author} {\bibinfo {author} {\bibfnamefont {R.}~\bibnamefont {Aguilera-Miret}}, \bibinfo {author} {\bibfnamefont {D.}~\bibnamefont {Vigan\`o}}, \bibinfo {author} {\bibfnamefont {F.}~\bibnamefont {Carrasco}}, \bibinfo {author} {\bibfnamefont {B.}~\bibnamefont {Mi\~nano}}, \ and\ \bibinfo {author} {\bibfnamefont {C.}~\bibnamefont {Palenzuela}},\ }\href {\doibase 10.1103/PhysRevD.102.103006} {\bibfield  {journal} {\bibinfo  {journal} {Phys. Rev. D}\ }\textbf {\bibinfo {volume} {102}},\ \bibinfo {pages} {103006} (\bibinfo {year} {2020})},\ \Eprint {http://arxiv.org/abs/2009.06669} {arXiv:2009.06669 [gr-qc]} \BibitemShut {NoStop}%
\bibitem [{\citenamefont {Balbus}\ and\ \citenamefont {Hawley}(1991)}]{Balbus:1991ay}%
  \BibitemOpen
  \bibfield  {author} {\bibinfo {author} {\bibfnamefont {S.~A.}\ \bibnamefont {Balbus}}\ and\ \bibinfo {author} {\bibfnamefont {J.~F.}\ \bibnamefont {Hawley}},\ }\href {\doibase 10.1086/170270} {\bibfield  {journal} {\bibinfo  {journal} {Astrophys. J.}\ }\textbf {\bibinfo {volume} {376}},\ \bibinfo {pages} {214} (\bibinfo {year} {1991})}\BibitemShut {NoStop}%
\bibitem [{\citenamefont {Shapiro}(2000)}]{Shapiro:2000zh}%
  \BibitemOpen
  \bibfield  {author} {\bibinfo {author} {\bibfnamefont {S.~L.}\ \bibnamefont {Shapiro}},\ }\href {\doibase 10.1086/317209} {\bibfield  {journal} {\bibinfo  {journal} {Astrophys. J.}\ }\textbf {\bibinfo {volume} {544}},\ \bibinfo {pages} {397} (\bibinfo {year} {2000})},\ \Eprint {http://arxiv.org/abs/astro-ph/0010493} {arXiv:astro-ph/0010493} \BibitemShut {NoStop}%
\bibitem [{\citenamefont {Duez}\ \emph {et~al.}(2006)\citenamefont {Duez}, \citenamefont {Liu}, \citenamefont {Shapiro}, \citenamefont {Shibata},\ and\ \citenamefont {Stephens}}]{Duez:2006qe}%
  \BibitemOpen
  \bibfield  {author} {\bibinfo {author} {\bibfnamefont {M.~D.}\ \bibnamefont {Duez}}, \bibinfo {author} {\bibfnamefont {Y.~T.}\ \bibnamefont {Liu}}, \bibinfo {author} {\bibfnamefont {S.~L.}\ \bibnamefont {Shapiro}}, \bibinfo {author} {\bibfnamefont {M.}~\bibnamefont {Shibata}}, \ and\ \bibinfo {author} {\bibfnamefont {B.~C.}\ \bibnamefont {Stephens}},\ }\href {\doibase 10.1103/PhysRevD.73.104015} {\bibfield  {journal} {\bibinfo  {journal} {Phys. Rev. D}\ }\textbf {\bibinfo {volume} {73}},\ \bibinfo {pages} {104015} (\bibinfo {year} {2006})},\ \Eprint {http://arxiv.org/abs/astro-ph/0605331} {arXiv:astro-ph/0605331} \BibitemShut {NoStop}%
\bibitem [{\citenamefont {Kastaun}\ and\ \citenamefont {Galeazzi}(2015)}]{Kastaun:2014fna}%
  \BibitemOpen
  \bibfield  {author} {\bibinfo {author} {\bibfnamefont {W.}~\bibnamefont {Kastaun}}\ and\ \bibinfo {author} {\bibfnamefont {F.}~\bibnamefont {Galeazzi}},\ }\href {\doibase 10.1103/PhysRevD.91.064027} {\bibfield  {journal} {\bibinfo  {journal} {Phys. Rev. D}\ }\textbf {\bibinfo {volume} {91}},\ \bibinfo {pages} {064027} (\bibinfo {year} {2015})},\ \Eprint {http://arxiv.org/abs/1411.7975} {arXiv:1411.7975 [gr-qc]} \BibitemShut {NoStop}%
\bibitem [{\citenamefont {Hanauske}\ \emph {et~al.}(2017)\citenamefont {Hanauske}, \citenamefont {Takami}, \citenamefont {Bovard}, \citenamefont {Rezzolla}, \citenamefont {Font}, \citenamefont {Galeazzi},\ and\ \citenamefont {St\"ocker}}]{Hanauske:2016gia}%
  \BibitemOpen
  \bibfield  {author} {\bibinfo {author} {\bibfnamefont {M.}~\bibnamefont {Hanauske}}, \bibinfo {author} {\bibfnamefont {K.}~\bibnamefont {Takami}}, \bibinfo {author} {\bibfnamefont {L.}~\bibnamefont {Bovard}}, \bibinfo {author} {\bibfnamefont {L.}~\bibnamefont {Rezzolla}}, \bibinfo {author} {\bibfnamefont {J.~A.}\ \bibnamefont {Font}}, \bibinfo {author} {\bibfnamefont {F.}~\bibnamefont {Galeazzi}}, \ and\ \bibinfo {author} {\bibfnamefont {H.}~\bibnamefont {St\"ocker}},\ }\href {\doibase 10.1103/PhysRevD.96.043004} {\bibfield  {journal} {\bibinfo  {journal} {Phys. Rev. D}\ }\textbf {\bibinfo {volume} {96}},\ \bibinfo {pages} {043004} (\bibinfo {year} {2017})},\ \Eprint {http://arxiv.org/abs/1611.07152} {arXiv:1611.07152 [gr-qc]} \BibitemShut {NoStop}%
\bibitem [{\citenamefont {Most}\ \emph {et~al.}(2019)\citenamefont {Most}, \citenamefont {Papenfort}, \citenamefont {Dexheimer}, \citenamefont {Hanauske}, \citenamefont {Schramm}, \citenamefont {St\"ocker},\ and\ \citenamefont {Rezzolla}}]{Most:2018eaw}%
  \BibitemOpen
  \bibfield  {author} {\bibinfo {author} {\bibfnamefont {E.~R.}\ \bibnamefont {Most}}, \bibinfo {author} {\bibfnamefont {L.~J.}\ \bibnamefont {Papenfort}}, \bibinfo {author} {\bibfnamefont {V.}~\bibnamefont {Dexheimer}}, \bibinfo {author} {\bibfnamefont {M.}~\bibnamefont {Hanauske}}, \bibinfo {author} {\bibfnamefont {S.}~\bibnamefont {Schramm}}, \bibinfo {author} {\bibfnamefont {H.}~\bibnamefont {St\"ocker}}, \ and\ \bibinfo {author} {\bibfnamefont {L.}~\bibnamefont {Rezzolla}},\ }\href {\doibase 10.1103/PhysRevLett.122.061101} {\bibfield  {journal} {\bibinfo  {journal} {Phys. Rev. Lett.}\ }\textbf {\bibinfo {volume} {122}},\ \bibinfo {pages} {061101} (\bibinfo {year} {2019})},\ \Eprint {http://arxiv.org/abs/1807.03684} {arXiv:1807.03684 [astro-ph.HE]} \BibitemShut {NoStop}%
\bibitem [{\citenamefont {Rivieccio}\ \emph {et~al.}(2024)\citenamefont {Rivieccio}, \citenamefont {Guerra}, \citenamefont {Ruiz},\ and\ \citenamefont {Font}}]{Rivieccio:2024sfm}%
  \BibitemOpen
  \bibfield  {author} {\bibinfo {author} {\bibfnamefont {G.}~\bibnamefont {Rivieccio}}, \bibinfo {author} {\bibfnamefont {D.}~\bibnamefont {Guerra}}, \bibinfo {author} {\bibfnamefont {M.}~\bibnamefont {Ruiz}}, \ and\ \bibinfo {author} {\bibfnamefont {J.~A.}\ \bibnamefont {Font}},\ }\href {\doibase 10.1103/PhysRevD.109.064032} {\bibfield  {journal} {\bibinfo  {journal} {Phys. Rev. D}\ }\textbf {\bibinfo {volume} {109}},\ \bibinfo {pages} {064032} (\bibinfo {year} {2024})},\ \Eprint {http://arxiv.org/abs/2401.06849} {arXiv:2401.06849 [astro-ph.HE]} \BibitemShut {NoStop}%
\bibitem [{\citenamefont {Raithel}\ and\ \citenamefont {Paschalidis}(2024)}]{Raithel:2023gct}%
  \BibitemOpen
  \bibfield  {author} {\bibinfo {author} {\bibfnamefont {C.~A.}\ \bibnamefont {Raithel}}\ and\ \bibinfo {author} {\bibfnamefont {V.}~\bibnamefont {Paschalidis}},\ }\href {\doibase 10.1103/PhysRevD.110.043002} {\bibfield  {journal} {\bibinfo  {journal} {Phys. Rev. D}\ }\textbf {\bibinfo {volume} {110}},\ \bibinfo {pages} {043002} (\bibinfo {year} {2024})},\ \Eprint {http://arxiv.org/abs/2312.14046} {arXiv:2312.14046 [astro-ph.HE]} \BibitemShut {NoStop}%
\bibitem [{\citenamefont {Fields}\ \emph {et~al.}(2023)\citenamefont {Fields}, \citenamefont {Prakash}, \citenamefont {Breschi}, \citenamefont {Radice}, \citenamefont {Bernuzzi},\ and\ \citenamefont {da~Silva~Schneider}}]{Fields2023}%
  \BibitemOpen
  \bibfield  {author} {\bibinfo {author} {\bibfnamefont {J.}~\bibnamefont {Fields}}, \bibinfo {author} {\bibfnamefont {A.}~\bibnamefont {Prakash}}, \bibinfo {author} {\bibfnamefont {M.}~\bibnamefont {Breschi}}, \bibinfo {author} {\bibfnamefont {D.}~\bibnamefont {Radice}}, \bibinfo {author} {\bibfnamefont {S.}~\bibnamefont {Bernuzzi}}, \ and\ \bibinfo {author} {\bibfnamefont {A.}~\bibnamefont {da~Silva~Schneider}},\ }\href {\doibase 10.3847/2041-8213/ace5b2} {\bibfield  {journal} {\bibinfo  {journal} {The Astrophysical Journal Letters}\ }\textbf {\bibinfo {volume} {952}},\ \bibinfo {pages} {L36} (\bibinfo {year} {2023})}\BibitemShut {NoStop}%
\bibitem [{\citenamefont {Villa-Ortega}\ \emph {et~al.}(2023)\citenamefont {Villa-Ortega}, \citenamefont {Lorenzo-Medina}, \citenamefont {Calder\'on~Bustillo}, \citenamefont {Ruiz}, \citenamefont {Guerra}, \citenamefont {Cerd\'a-Duran},\ and\ \citenamefont {Font}}]{Villa-Ortega:2023cps}%
  \BibitemOpen
  \bibfield  {author} {\bibinfo {author} {\bibfnamefont {V.}~\bibnamefont {Villa-Ortega}}, \bibinfo {author} {\bibfnamefont {A.}~\bibnamefont {Lorenzo-Medina}}, \bibinfo {author} {\bibfnamefont {J.}~\bibnamefont {Calder\'on~Bustillo}}, \bibinfo {author} {\bibfnamefont {M.}~\bibnamefont {Ruiz}}, \bibinfo {author} {\bibfnamefont {D.}~\bibnamefont {Guerra}}, \bibinfo {author} {\bibfnamefont {P.}~\bibnamefont {Cerd\'a-Duran}}, \ and\ \bibinfo {author} {\bibfnamefont {J.~A.}\ \bibnamefont {Font}},\ }\href@noop {} {\  (\bibinfo {year} {2023})},\ \Eprint {http://arxiv.org/abs/2310.20378} {arXiv:2310.20378 [gr-qc]} \BibitemShut {NoStop}%
\bibitem [{\citenamefont {Raithel}\ and\ \citenamefont {Most}(2022)}]{Raithel:2022orm}%
  \BibitemOpen
  \bibfield  {author} {\bibinfo {author} {\bibfnamefont {C.~A.}\ \bibnamefont {Raithel}}\ and\ \bibinfo {author} {\bibfnamefont {E.~R.}\ \bibnamefont {Most}},\ }\href {\doibase 10.3847/2041-8213/ac7c75} {\bibfield  {journal} {\bibinfo  {journal} {Astrophys. J. Lett.}\ }\textbf {\bibinfo {volume} {933}},\ \bibinfo {pages} {L39} (\bibinfo {year} {2022})},\ \Eprint {http://arxiv.org/abs/2201.03594} {arXiv:2201.03594 [astro-ph.HE]} \BibitemShut {NoStop}%
\bibitem [{\citenamefont {Chabanov}\ and\ \citenamefont {Rezzolla}(2023)}]{Chabanov:2023blf}%
  \BibitemOpen
  \bibfield  {author} {\bibinfo {author} {\bibfnamefont {M.}~\bibnamefont {Chabanov}}\ and\ \bibinfo {author} {\bibfnamefont {L.}~\bibnamefont {Rezzolla}},\ }\href@noop {} {\  (\bibinfo {year} {2023})},\ \Eprint {http://arxiv.org/abs/2307.10464} {arXiv:2307.10464 [gr-qc]} \BibitemShut {NoStop}%
\bibitem [{\citenamefont {Most}\ \emph {et~al.}(2024)\citenamefont {Most}, \citenamefont {Haber}, \citenamefont {Harris}, \citenamefont {Zhang}, \citenamefont {Alford},\ and\ \citenamefont {Noronha}}]{Most_2024}%
  \BibitemOpen
  \bibfield  {author} {\bibinfo {author} {\bibfnamefont {E.~R.}\ \bibnamefont {Most}}, \bibinfo {author} {\bibfnamefont {A.}~\bibnamefont {Haber}}, \bibinfo {author} {\bibfnamefont {S.~P.}\ \bibnamefont {Harris}}, \bibinfo {author} {\bibfnamefont {Z.}~\bibnamefont {Zhang}}, \bibinfo {author} {\bibfnamefont {M.~G.}\ \bibnamefont {Alford}}, \ and\ \bibinfo {author} {\bibfnamefont {J.}~\bibnamefont {Noronha}},\ }\href {\doibase 10.3847/2041-8213/ad454f} {\bibfield  {journal} {\bibinfo  {journal} {The Astrophysical Journal Letters}\ }\textbf {\bibinfo {volume} {967}},\ \bibinfo {pages} {L14} (\bibinfo {year} {2024})}\BibitemShut {NoStop}%
\bibitem [{\citenamefont {East}\ \emph {et~al.}(2019)\citenamefont {East}, \citenamefont {Paschalidis}, \citenamefont {Pretorius},\ and\ \citenamefont {Tsokaros}}]{East:2019lbk}%
  \BibitemOpen
  \bibfield  {author} {\bibinfo {author} {\bibfnamefont {W.~E.}\ \bibnamefont {East}}, \bibinfo {author} {\bibfnamefont {V.}~\bibnamefont {Paschalidis}}, \bibinfo {author} {\bibfnamefont {F.}~\bibnamefont {Pretorius}}, \ and\ \bibinfo {author} {\bibfnamefont {A.}~\bibnamefont {Tsokaros}},\ }\href {\doibase 10.1103/PhysRevD.100.124042} {\bibfield  {journal} {\bibinfo  {journal} {Phys. Rev. D}\ }\textbf {\bibinfo {volume} {100}},\ \bibinfo {pages} {124042} (\bibinfo {year} {2019})},\ \Eprint {http://arxiv.org/abs/1906.05288} {arXiv:1906.05288 [astro-ph.HE]} \BibitemShut {NoStop}%
\bibitem [{\citenamefont {Lam}\ \emph {et~al.}(2024)\citenamefont {Lam}, \citenamefont {Gao}, \citenamefont {Kuan}, \citenamefont {Shibata}, \citenamefont {Van~Aelst},\ and\ \citenamefont {Kiuchi}}]{Lam:2024azd}%
  \BibitemOpen
  \bibfield  {author} {\bibinfo {author} {\bibfnamefont {A.~T.-L.}\ \bibnamefont {Lam}}, \bibinfo {author} {\bibfnamefont {Y.}~\bibnamefont {Gao}}, \bibinfo {author} {\bibfnamefont {H.-J.}\ \bibnamefont {Kuan}}, \bibinfo {author} {\bibfnamefont {M.}~\bibnamefont {Shibata}}, \bibinfo {author} {\bibfnamefont {K.}~\bibnamefont {Van~Aelst}}, \ and\ \bibinfo {author} {\bibfnamefont {K.}~\bibnamefont {Kiuchi}},\ }\href@noop {} {\  (\bibinfo {year} {2024})},\ \Eprint {http://arxiv.org/abs/2410.00137} {arXiv:2410.00137 [astro-ph.HE]} \BibitemShut {NoStop}%
\bibitem [{\citenamefont {Andersson}\ and\ \citenamefont {Kokkotas}(1998)}]{Andersson:1997rn}%
  \BibitemOpen
  \bibfield  {author} {\bibinfo {author} {\bibfnamefont {N.}~\bibnamefont {Andersson}}\ and\ \bibinfo {author} {\bibfnamefont {K.~D.}\ \bibnamefont {Kokkotas}},\ }\href {\doibase 10.1046/j.1365-8711.1998.01840.x} {\bibfield  {journal} {\bibinfo  {journal} {Mon. Not. Roy. Astron. Soc.}\ }\textbf {\bibinfo {volume} {299}},\ \bibinfo {pages} {1059} (\bibinfo {year} {1998})},\ \Eprint {http://arxiv.org/abs/gr-qc/9711088} {arXiv:gr-qc/9711088} \BibitemShut {NoStop}%
\bibitem [{\citenamefont {Chakravarti}\ and\ \citenamefont {Andersson}(2020)}]{Chakravarti:2019sdc}%
  \BibitemOpen
  \bibfield  {author} {\bibinfo {author} {\bibfnamefont {K.}~\bibnamefont {Chakravarti}}\ and\ \bibinfo {author} {\bibfnamefont {N.}~\bibnamefont {Andersson}},\ }\href {\doibase 10.1093/mnras/staa2342} {\bibfield  {journal} {\bibinfo  {journal} {Mon. Not. Roy. Astron. Soc.}\ }\textbf {\bibinfo {volume} {497}},\ \bibinfo {pages} {5480} (\bibinfo {year} {2020})},\ \Eprint {http://arxiv.org/abs/1906.04546} {arXiv:1906.04546 [gr-qc]} \BibitemShut {NoStop}%
\bibitem [{\citenamefont {Sekiguchi}\ \emph {et~al.}(2011{\natexlab{b}})\citenamefont {Sekiguchi}, \citenamefont {Kiuchi}, \citenamefont {Kyutoku},\ and\ \citenamefont {Shibata}}]{Sekiguchi2011}%
  \BibitemOpen
  \bibfield  {author} {\bibinfo {author} {\bibfnamefont {Y.}~\bibnamefont {Sekiguchi}}, \bibinfo {author} {\bibfnamefont {K.}~\bibnamefont {Kiuchi}}, \bibinfo {author} {\bibfnamefont {K.}~\bibnamefont {Kyutoku}}, \ and\ \bibinfo {author} {\bibfnamefont {M.}~\bibnamefont {Shibata}},\ }\href {\doibase 10.1103/PhysRevLett.107.211101} {\bibfield  {journal} {\bibinfo  {journal} {Phys. Rev. Lett.}\ }\textbf {\bibinfo {volume} {107}},\ \bibinfo {pages} {211101} (\bibinfo {year} {2011}{\natexlab{b}})}\BibitemShut {NoStop}%
\bibitem [{\citenamefont {Hotokezaka}\ \emph {et~al.}(2013)\citenamefont {Hotokezaka}, \citenamefont {Kiuchi}, \citenamefont {Kyutoku}, \citenamefont {Muranushi}, \citenamefont {Sekiguchi}, \citenamefont {Shibata},\ and\ \citenamefont {Taniguchi}}]{Hotokezaka2013}%
  \BibitemOpen
  \bibfield  {author} {\bibinfo {author} {\bibfnamefont {K.}~\bibnamefont {Hotokezaka}}, \bibinfo {author} {\bibfnamefont {K.}~\bibnamefont {Kiuchi}}, \bibinfo {author} {\bibfnamefont {K.}~\bibnamefont {Kyutoku}}, \bibinfo {author} {\bibfnamefont {T.}~\bibnamefont {Muranushi}}, \bibinfo {author} {\bibfnamefont {Y.-i.}\ \bibnamefont {Sekiguchi}}, \bibinfo {author} {\bibfnamefont {M.}~\bibnamefont {Shibata}}, \ and\ \bibinfo {author} {\bibfnamefont {K.}~\bibnamefont {Taniguchi}},\ }\href {\doibase 10.1103/PhysRevD.88.044026} {\bibfield  {journal} {\bibinfo  {journal} {Phys. Rev. D}\ }\textbf {\bibinfo {volume} {88}},\ \bibinfo {pages} {044026} (\bibinfo {year} {2013})}\BibitemShut {NoStop}%
\bibitem [{\citenamefont {Radice}\ \emph {et~al.}(2017)\citenamefont {Radice}, \citenamefont {Bernuzzi}, \citenamefont {Del~Pozzo}, \citenamefont {Roberts},\ and\ \citenamefont {Ott}}]{Radice:2016rys}%
  \BibitemOpen
  \bibfield  {author} {\bibinfo {author} {\bibfnamefont {D.}~\bibnamefont {Radice}}, \bibinfo {author} {\bibfnamefont {S.}~\bibnamefont {Bernuzzi}}, \bibinfo {author} {\bibfnamefont {W.}~\bibnamefont {Del~Pozzo}}, \bibinfo {author} {\bibfnamefont {L.~F.}\ \bibnamefont {Roberts}}, \ and\ \bibinfo {author} {\bibfnamefont {C.~D.}\ \bibnamefont {Ott}},\ }\href {\doibase 10.3847/2041-8213/aa775f} {\bibfield  {journal} {\bibinfo  {journal} {Astrophys. J. Lett.}\ }\textbf {\bibinfo {volume} {842}},\ \bibinfo {pages} {L10} (\bibinfo {year} {2017})},\ \Eprint {http://arxiv.org/abs/1612.06429} {arXiv:1612.06429 [astro-ph.HE]} \BibitemShut {NoStop}%
\bibitem [{\citenamefont {Chatziioannou}\ \emph {et~al.}(2017)\citenamefont {Chatziioannou}, \citenamefont {Clark}, \citenamefont {Bauswein}, \citenamefont {Millhouse}, \citenamefont {Littenberg},\ and\ \citenamefont {Cornish}}]{Chatziioannou2017}%
  \BibitemOpen
  \bibfield  {author} {\bibinfo {author} {\bibfnamefont {K.}~\bibnamefont {Chatziioannou}}, \bibinfo {author} {\bibfnamefont {J.~A.}\ \bibnamefont {Clark}}, \bibinfo {author} {\bibfnamefont {A.}~\bibnamefont {Bauswein}}, \bibinfo {author} {\bibfnamefont {M.}~\bibnamefont {Millhouse}}, \bibinfo {author} {\bibfnamefont {T.~B.}\ \bibnamefont {Littenberg}}, \ and\ \bibinfo {author} {\bibfnamefont {N.}~\bibnamefont {Cornish}},\ }\href {\doibase 10.1103/PhysRevD.96.124035} {\bibfield  {journal} {\bibinfo  {journal} {Phys. Rev. D}\ }\textbf {\bibinfo {volume} {96}},\ \bibinfo {pages} {124035} (\bibinfo {year} {2017})}\BibitemShut {NoStop}%
\bibitem [{\citenamefont {Bauswein}\ \emph {et~al.}(2014)\citenamefont {Bauswein}, \citenamefont {Stergioulas},\ and\ \citenamefont {Janka}}]{Bauswein:2014qla}%
  \BibitemOpen
  \bibfield  {author} {\bibinfo {author} {\bibfnamefont {A.}~\bibnamefont {Bauswein}}, \bibinfo {author} {\bibfnamefont {N.}~\bibnamefont {Stergioulas}}, \ and\ \bibinfo {author} {\bibfnamefont {H.~T.}\ \bibnamefont {Janka}},\ }\href {\doibase 10.1103/PhysRevD.90.023002} {\bibfield  {journal} {\bibinfo  {journal} {Phys. Rev. D}\ }\textbf {\bibinfo {volume} {90}},\ \bibinfo {pages} {023002} (\bibinfo {year} {2014})},\ \Eprint {http://arxiv.org/abs/1403.5301} {arXiv:1403.5301 [astro-ph.SR]} \BibitemShut {NoStop}%
\end{thebibliography}%

\end{document}